\pdfoutput=0
\documentclass[11pt]{article}
\usepackage{axodraw}
\usepackage{epsfig}
\usepackage{amsfonts}
\usepackage{amsmath}
\usepackage{bm,bbm}
\usepackage{cite}
\usepackage{hyperref}
\allowdisplaybreaks[1]

\input pix.sty

\newcommand{\la}[1]{\label{#1}}
\newcommand{\be}{\begin{equation}}
\newcommand{\ee}{\end{equation}}
\newcommand{\ba}{\begin{eqnarray}}
\newcommand{\ea}{\end{eqnarray}}
\newcommand{\rmi}[1]{{\mbox{\scriptsize #1}}}
\newcommand{\fig}{Fig.~}

\newcommand{\eq}{Eq.~}
\newcommand{\eqs}{Eqs.~}
\newcommand{\se}{Sec.~}
\newcommand{\ses}{Secs.~}
\newcommand{\nr}[1]{(\ref{#1})}

\newcommand{\nn}{\nonumber \\}
\newcommand{\fr}[2]{{\frac{#1}{#2}\,}}

\renewcommand{\vec}[1]{{\bf #1}}

\newcommand{\tfr}[2]{{\textstyle \frac{#1}{#2}\,}}

\renewcommand{\eq}{eq.~}
\renewcommand{\eqs}{eqs.~}
\renewcommand{\se}{sec.~}
\renewcommand{\ses}{secs.~}
\renewcommand{\fig}{fig.~}

\newcommand{\PT}{\mathbbm{P}^\rmii{T}}
\newcommand{\PE}{\mathbbm{P}^\rmii{E}}

\newcommand{\alphas}{\alpha_{\rm s}}
\renewcommand{\B}{\rmii{$B$}}

\newcommand{\T}{\rmii{$T$}}

\newcommand{\mpl}{m_\rmi{pl}}

\newcommand{\Nc}{N_{\rm c}}

\newcommand{\Tc}{T_{\rm c}}
\newcommand{\nG}{n_\rmii{G}}
\newcommand{\nS}{n_\rmii{S}}

\newcommand{\mE}{m_\rmii{E2}}
\newcommand{\mEa}{m_\rmii{E$a$}}
\newcommand{\mF}{m_\rmii{F}}
\newcommand{\mFmini}{m_\rmiii{F}}

\newcommand{\rmO}{{\mathcal{O}}}

\newcommand{\CF}{C_\rmii{F}}

\def\lsi{\raise0.3ex\hbox{$<$\kern-0.75em\raise-1.1ex\hbox{$\sim$}}}
\def\gsi{\raise0.3ex\hbox{$>$\kern-0.75em\raise-1.1ex\hbox{$\sim$}}}
\newcommand{\lsim}{\mathop{\lsi}}
\newcommand{\gsim}{\mathop{\gsi}}

\newcommand{\sign}{\mathop{\mbox{sign}}}
\newcommand{\nF}{n_\rmii{F}} 
\newcommand{\nB}{n_\rmii{B}} 
\newcommand{\rmii}[1]{{\mbox{\tiny\rm{#1}}}}
\newcommand{\rmiii}[1]{{\mbox{\tiny{$\scriptstyle{\rm#1}$}}}}
\newcommand{\re}{\mathop{\mbox{Re}}}
\newcommand{\im}{\mathop{\mbox{Im}}}

\newcommand{\Tint}[1]{{\hbox{$\sum$}\!\!\!\!\!\!\!\int\,}_{\!\!\!\!\raise-0.9ex\hbox{$\scriptstyle{#1}$}}}
\newcommand{\Tinti}[1]{{{\Sigma}\!\!\!\!\raise0.3ex\hbox{$\int$}_\rmii{${#1}$}}}

\newcommand{\bi}{\begin{itemize}}
\newcommand{\ei}{\end{itemize}}
\newcommand{\hide}[1]{ }
\newcommand{\blind}[1]{\fbox{$ ? $}} 
\newcommand{\bsl}[1]{\,\slash\!\!\!\!{#1}\,}

\newcommand{\deltabar}{\raise-0.02em\hbox{$\bar{}$}\hspace*{-0.8mm}{\delta}}

\renewcommand{\P}{\mathcal{P}}

\newcommand{\K}{\mathcal{K}}
\newcommand{\X}{\mathcal{X}}
\newcommand{\Y}{\mathcal{Y}}

\newcommand{\mZ}{m_\rmii{$Z$}} 
\newcommand{\mW}{m_\rmii{$W$}} 
\newcommand{\frict}{\Upsilon} 
\newcommand{\channel}{\alpha} 
\def\TAsc(#1,#2)(#3,#4,#5)%
{\SetWidth{2.0}\CArc(#1,#2)(#3,#4,#5)\SetWidth{1.0}}
\def\TLsc(#1,#2)(#3,#4)%
{\SetWidth{2.0}\Line(#1,#2)(#3,#4)\SetWidth{1.0}}
\def\Lwidth{3}

\def\TAgl(#1,#2)(#3,#4,#5){\SetWidth{2.0}\PhotonArc(#1,#2)(#3,#4,#5){\Lwidth}%
{6.283 #3 mul 360 div #4 #5 sub #4 #5 sub mul sqrt mul Tdensity mul}%
\SetWidth{1.0}}
\def\TLgl(#1,#2)(#3,#4){\SetWidth{2.0}\Photon(#1,#2)(#3,#4){\Lwidth}
{#1 #3 sub #1 #3 sub mul #2 #4 sub #2 #4 sub mul add sqrt Tdensity mul}%
\SetWidth{1.0}}

\renewcommand{\picc}[1]{\;\parbox[c]{60pt}{\begin{picture}(60,30)(0,-10)
\SetWidth{1.0}\SetScale{1.3} #1 \end{picture}}\; }
\def\Lwidth{1.3}

\renewcommand{\pic}[1]{\;\parbox[c]{30pt}{\begin{picture}(30,30)(0,-3)
\SetWidth{1.0}\SetScale{0.8} #1 \end{picture}}\;}
\renewcommand{\picb}[1]{\;\parbox[c]{45pt}{\begin{picture}(45,30)(0,-3)
\SetWidth{1.0}\SetScale{0.8} #1 \end{picture}}\;}
\def\cutC{\picb{%
 \Agl(25,15)(15,0,180)%
 \Asc(25,15)(15,180,360)%
 \Lsc(0,15)(10,15)%
 \Lsc(40,15)(50,15)%
 \CCirc(25,30){3}{Black}{Black}%
 \Line(25,-8)(25,38)
 \Line(20,-10)(25,-8)
 \Line(25,38)(30,40)
}}
\def\cutF{\picb{%
 \Aqu(25,15)(15,0,100)%
 \CArc(25,15)(15,100,180)%
 \Aqu(25,15)(15,180,280)%
 \CArc(25,15)(15,280,360)%
 \Lsc(0,15)(10,15)%
 \Lsc(40,15)(50,15)%
 \CCirc(25,0){3}{Black}{Black}%
 \CCirc(25,30){3}{Black}{Black}%
 \Line(25,-8)(25,38)
 \Line(20,-10)(25,-8)
 \Line(25,38)(30,40)
}}
\def\cutHa{\picb{%
 \Asc(25,15)(15,0,180)%
 \Asc(25,15)(15,180,360)%
 \Lsc(0,15)(10,15)%
 \Lsc(40,15)(50,15)%
 \Line(25,-8)(25,38)
 \Line(20,-10)(25,-8)
 \Line(25,38)(30,40)
}}
\def\cutHb{\picb{%
 \Agl(25,15)(15,0,180)%
 \Asc(25,15)(15,180,360)%
 \Lsc(0,15)(10,15)%
 \Lsc(40,15)(50,15)%
 \CCirc(25,30){3}{Black}{Black}%
 \Line(25,-8)(25,38)
 \Line(20,-10)(25,-8)
 \Line(25,38)(30,40)
}}
\def\cutHc{\picb{%
 \Agl(25,15)(15,0,180)%
 \Agl(25,15)(15,180,360)%
 \Lsc(0,15)(10,15)%
 \Lsc(40,15)(50,15)%
 \CCirc(25,0){3}{Black}{Black}%
 \CCirc(25,30){3}{Black}{Black}%
 \Line(25,-8)(25,38)
 \Line(20,-10)(25,-8)
 \Line(25,38)(30,40)
}}
\def\cutHd{\picb{%
 \Agh(25,15)(15,0,180)%
 \Agh(25,15)(15,180,360)%
 \Lsc(0,15)(10,15)%
 \Lsc(40,15)(50,15)%
 \Line(25,-8)(25,38)
 \Line(20,-10)(25,-8)
 \Line(25,38)(30,40)
}}
\def\ampA{\picb{%
 \Lgl(22,15)(50,15)%
 \Lsc(0,0)(22,15)%
 \Lsc(0,30)(22,15)%
 \CCirc(35,15){3}{Black}{Black}%
}}
\def\ampTmini{\pic{%
 \Lgl(25,5)(25,30)%
 \Lsc(0,0)(25,5)%
 \Lsc(25,5)(50,0)%
 \CCirc(25,15){3}{Black}{Black}%
 \GOval(25,30)(15,5)(90){0.5}
 \Text(0,6)[c]{$\scriptstyle \omega$}
 \Text(43,6)[c]{$\scriptstyle \epsilon^{ }_\phi$}
 \Text(24,13)[l]{$\scriptstyle p^0_{ }$}
}}
\def\ampSmini{\pic{%
 \Lgl(22,15)(50,15)%
 \Lsc(0,0)(22,15)%
 \Lsc(0,30)(22,15)%
 \CCirc(33,15){3}{Black}{Black}%
 \GOval(50,15)(15,5)(0){0.5}
 \Text(-6,24)[c]{$\scriptstyle \omega$}
 \Text(-6,-2)[c]{$\scriptstyle \epsilon^{ }_\phi$}
 \Text(24,5)[l]{$\scriptstyle p^0_{ }$}
}}
\def\ampC{\picb{%
 \Lgl(25,5)(25,25)%
 \Lsc(0,0)(25,5)%
 \Lsc(25,5)(50,0)%
 \CCirc(25,15){3}{Black}{Black}%
 \Line(0,30)(25,25)
 \Line(25,25)(50,30)
 \LongArrow(18,10)(18,20)
 \LongArrow(0,5)(10,7)
 \LongArrow(40,7)(50,5)
 \Text(0,-8)[c]{$\scriptstyle (\omega,\vec{0})$}
 \Text(50,-8)[c]{$\scriptstyle (\epsilon^{ }_\phi,\vec{p})$}
 \Text(26,12)[l]{$\scriptstyle (\omega-\epsilon^{ }_\phi,-\vec{p})$}
}}
\def\ampFa{\picb{%
 \Lsc(0,0)(25,5)%
 \TLsc(25,5)(50,0)%
 \CCirc(37.5,2.5){3}{Black}{Black}%
 \TLsc(25,5)(25,25)%
 \CCirc(25,15){3}{Black}{Black}%
 \TLsc(0,30)(25,25)
 \Gluon(25,25)(50,30){3}{3} 
}}
\def\ampFb{\picb{%
 \Lsc(0,0)(25,5)%
 \TLsc(25,5)(25,25)%
 \CCirc(25,15){3}{Black}{Black}%
 \TLsc(25,5)(50,0)%
 \CCirc(37.5,2.5){3}{Black}{Black}%
 \TLsc(50,30)(25,25)
 \Gluon(25,25)(0,30){-3}{3} 
}}
\def\ampFFa{\picb{%
 \Lsc(-5,15)(25,15)%
 \TLsc(25,15)(25,35)%
 \CCirc(25,25){3}{Black}{Black}%
 \TLsc(0,40)(25,35)
 \Gluon(25,35)(50,40){3}{3} 
 \TLsc(25,15)(25,-5)%
 \CCirc(25,5){3}{Black}{Black}%
 \TLsc(0,-10)(25,-5)
 \Gluon(25,-5)(50,-10){-3}{3} 
}}
\def\ampFFb{\picb{%
 \Lsc(-5,15)(25,15)%
 \TLsc(25,15)(25,35)%
 \CCirc(25,25){3}{Black}{Black}%
 \TLsc(0,40)(25,35)
 \Gluon(25,35)(50,40){3}{3} 
 \TLsc(25,15)(25,-5)%
 \CCirc(25,5){3}{Black}{Black}%
 \TLsc(50,-10)(25,-5)
 \Gluon(25,-5)(0,-10){3}{3} 
}}
\def\ampFFc{\picb{%
 \Lsc(-5,15)(25,15)%
 \TLsc(25,15)(25,35)%
 \CCirc(25,25){3}{Black}{Black}%
 \TLsc(50,40)(25,35)
 \Gluon(25,35)(0,40){-3}{3} 
 \TLsc(25,15)(25,-5)%
 \CCirc(25,5){3}{Black}{Black}%
 \TLsc(0,-10)(25,-5)
 \Gluon(25,-5)(50,-10){-3}{3} 
}}
\def\ampFFd{\picb{%
 \Lsc(-5,15)(25,15)%
 \TLsc(25,15)(25,35)%
 \CCirc(25,25){3}{Black}{Black}%
 \TLsc(50,40)(25,35)
 \Gluon(25,35)(0,40){-3}{3} 
 \TLsc(25,15)(25,-5)%
 \CCirc(25,5){3}{Black}{Black}%
 \TLsc(50,-10)(25,-5)
 \Gluon(25,-5)(0,-10){3}{3} 
}}
\def\ampBa{\picb{%
 \Lsc(0,0)(25,5)%
 \Lgl(25,5)(25,25)%
 \CCirc(25,15){3}{Black}{Black}%
 \Lgl(25,5)(50,0)%
 \CCirc(37.5,2.5){3}{Black}{Black}%
 \Line(0,30)(25,25)
 \Line(25,25)(50,30)
}}
\def\ampBBa{\picb{%
 \Lsc(-5,15)(25,15)%
 \Lgl(25,15)(25,35)%
 \CCirc(25,25){3}{Black}{Black}%
 \Line(50,40)(25,35)
 \Line(25,35)(0,40)
 \Lgl(25,15)(25,-5)%
 \CCirc(25,5){3}{Black}{Black}%
 \Line(50,-10)(25,-5)
 \Line(25,-5)(0,-10) 
}}

\makeatletter \@addtoreset{equation}{section} \makeatother
\renewcommand{\theequation}{\arabic{section}.\arabic{equation}}
\makeatletter
\renewcommand\section{\@startsection {section}{1}{\z@}%
                                   {-5.5ex \@plus -1ex \@minus -.2ex}
                                   {2.3ex \@plus.2ex}%
                                   {\normalfont\large\bfseries}}
\renewcommand\subsection{\@startsection{subsection}{2}{\z@}%
                                     {-3.25ex\@plus -1ex \@minus -.2ex}%
                                     {1.5ex \@plus .2ex}%
                                     {\normalfont\normalsize\bfseries}}
\renewcommand\thesection {\@arabic\c@section}
\renewcommand\thesubsection   {\thesection.\@arabic\c@subsection}
\renewcommand{\@seccntformat}[1]{%
\csname the#1\endcsname.\hspace{1.0em}}
\makeatother

\begin{document}

\flushbottom

\begin{titlepage}

\begin{flushright}
May 2024
\end{flushright}
\begin{centering}
\vfill

{\Large{\bf
  Soft contributions to the thermal Higgs width \\[3mm]
  across an electroweak phase transition
}} 

\vspace{0.8cm}

M.~Eriksson, 
M.~Laine 

\vspace{0.8cm}

{\em
AEC, 
Institute for Theoretical Physics, 
University of Bern, \\ 
Sidlerstrasse 5, CH-3012 Bern, Switzerland \\}

\vspace*{0.8cm}

\mbox{\bf Abstract}
 
\end{centering}

\vspace*{0.3cm}
 
\noindent
We estimate the equilibration rate of 
a nearly homogeneous Higgs field,
displaced from its ground state during the onset 
of an electroweak phase transition.
The computation is carried out with Hard Thermal Loop
resummed perturbation theory, and a significant 
part of the result originates from Bose-enhanced 
$t$-channel $2\leftrightarrow 2$ scatterings. 
The expression is shown to be IR finite and gauge independent. 
Possible applications to Langevin simulations of bubble nucleation 
are mentioned, and we also contrast with 
the friction affecting bubble growth.

\vfill


\end{titlepage}

\tableofcontents

%
\section{Introduction}
\la{se:intro}

The recently approved LISA gravitational wave interferometer 
will offer unprecedented sensitivity to detect a possible 
weak-scale cosmological phase transition~\cite{lisa}. Even though
there is no such phase transition in the Standard Model of particle
physics, many simple extensions do display first-order phase 
transitions. Therefore, it appears well-motivated to develop tools
for a systematic study of their dynamics.

Many different momentum and energy scales (or length and time scales)
play a role for the physics of a phase transition. The most macroscopic
features of a phase transition, such as growing bubbles and
the subsequent complicated fluid motion, are 
described by hydrodynamics. More microscopic features may be captured
by kinetic theory, which should also permit for the computation of 
the transport coefficients appearing in the hydrodynamic equations. 
Some quantities, like the equation of state, are sensitive
to the very smallest length scales, 
and need to be derived directly from the underlying quantum field theory.  

In the intermediate domain of kinetic theory, which describes 
what is frequently
referred to as ``soft'' physics, the relevant degrees of freedom are
quasiparticles, whose properties get modified by thermal 
corrections. The modifications are generally 
divided into two types: dispersive and absorptive effects.
Dispersive effects, such as thermal mass corrections, 
modify the relationship between the energy
and momentum of a quasiparticle; we may say that they shift 
the real part of a pole appearing in a propagator. 
Absorptive effects describe a ``width'', i.e.\ an interaction or
damping rate, and correspond to the imaginary part of the pole. 

A most prominent quasiparticle is the Higgs boson. 
In vacuum, it has a large mass, $m^{ }_{h,0} \approx 125$~GeV and, 
according to the Standard Model,  
a small width, $\Gamma^{ }_{h,0} \approx 4$~MeV. 

Both the Higgs mass and width change with the temperature.
The mass changes through the evolution of the vacuum expectation
value, originating from positive thermal corrections
$\sim \rmO(g^2T^2)$ to the Higgs mass parameter~\cite{meg}, 
where $g$ is a generic coupling constant. The width changes
due to the Bose enhancement or Pauli blocking experienced by the decay 
products of $1\to 2$ or $1\to 3$ processes, 
but also due to new reactions, such as $2\to 1$ ``inverse
decays'', as well as $2\to 2$
scatterings, involving additional particles in the
initial state~\cite{gw}. In addition, 
$1 + n \to 2+n$ and $2+n \to 1+n$ processes 
with $n\ge 1$, may play a role.   

\vspace*{3mm}

When we go to very high temperatures, $g T \sim m^{ }_{h,0}$, 
the positive thermal corrections to the Higgs mass parameter
cancel against the negative vacuum term, and the electroweak
symmetry is said to get restored. If
the system underwent a second order transition, 
the Higgs mass could be made arbitrarily small 
by approaching the phase transition point. 

In a more general situation, when the transition is a crossover
or of first order, the Higgs mass does not become arbitrarily small. 
Let us consider the energy of scalar excitations having 
a vanishing spatial momentum, 
and adopt it as a definition of a thermal Higgs mass, 
denoted by $m^{ }_h \equiv m^{ }_{h,\T}$. 
Different parametric regimes can be envisaged. 
Close to a crossover, 
perturbation theory breaks down, and we may say that the Higgs 
mass is of order $m^{ }_h \sim O(g^2T/\pi)$~\cite{linde}.
In contrast, in a first order transition, when the system interpolates
between two stable phases, the Higgs mass could be larger
in both of them (though it is tachyonic in between). 
For instance, assuming a partial cancellation
between the negative vacuum term $-m_{h,0}^2/2$ 
and the positive thermal correction $\sim\rmO(g^2T^2)$, 
we might assign the magnitude 
\be
 m^{2}_h \sim  \frac{ g^3 T^2 }{\pi} \ll g^2 T^2
 \la{regime}
\ee
to the effective thermal Higgs mass squared~\cite{gv2,lm}. 
Organizing perturbative computations
with this power counting has been argued 
to improve upon their convergence~\cite{gt}.


Under the assumption of \eq\nr{regime}, 
the Higgs boson is lighter than most other particles, whose thermal 
masses are $\sim \rmO(gT)$. Then its $1\to 2$ decays that dominate
in vacuum are no longer kinematically allowed. However, the Higgs
still has a thermal interaction rate. The purpose of this paper is to 
estimate this width, and show that the result is of order\footnote{%
 A similar result can be found already in ref.~\cite{htl_old}, 
 but we carry out a more comprehensive analysis.
 }
\be
 \Gamma^{ }_h \sim \frac{g^2 T}{\pi}
 \;. \la{result}
\ee

\vspace*{3mm}

It is an interesting and non-trivial question 
which physical role the Higgs width of \eq\nr{result} can play. 
We start this paper by briefly addressing 
this issue, in \se\ref{se:setup}. Then 
we proceed to the explicit 
computation of $\Gamma^{ }_h$, 
in \ses\ref{se:processes} (for $2\leftrightarrow 2$ processes)
and \ref{se:1n_2n} (for $1+n\leftrightarrow 2+n$ processes). 
The results are illustrated numerically in \se\ref{se:soft}, before
we turn to an outlook in \se\ref{se:concl}. Appendix~A 
complements the body of the text, 
in which scatterings involving weak gauge bosons are considered, 
by estimating at leading-logarithmic level 
the Yukawa-mediated contribution involving top quarks and gluons.

\section{Setup and motivation}
\la{se:setup}

A classic physics application for the electroweak phase transition 
is baryogenesis (for reviews see,\ e.g.,\ 
refs.~\cite{bary1,bary2}).
A key quantity in this context is the so-called sphaleron rate, 
proportional to the anomalous rate of baryon plus lepton number 
violation. The anomalous rate is generated by non-perturbative
gauge field dynamics, at the momentum scale 
$k  \equiv |\vec{k}| \sim g^2_{ }T/\pi$.
The gauge fields are the ``slow'' or ``ultrasoft'' modes 
of this problem, with a characteristic evolution rate
$\sim g^4_{ }\ln(1/g) T / \pi^3_{ }$~\cite{asy,db}. However, 
the Higgs doublet also affects the sphaleron dynamics~\cite{sph_gdm}.
Its evolution rate, from \eq\nr{result}, is just much faster. 
Therefore, in practical simulations~\cite{sph_sm}, 
the Higgs doublet can be 
assumed to thermalize within a given gauge background,
whereas  the gauge fields evolve slowly via first-order (overdamped) 
gauge-covariant Langevin equations~\cite{db}. 

The very same framework that is used for determining the sphaleron
rate, has also been applied to bubble nucleation~\cite{bbl1,bbl2}.
However, there is a qualitative difference between the sphaleron rate
and the bubble nucleation rate. 
While the sphaleron rate is a smooth function of the temperature, 
and determined by its local value within a radius comparable to 
a magnetic screening length,  
$\ell^{ }_\rmii{M} \sim \pi / (g^2_{ }T)$, 
the length scale that is relevant for bubble nucleations 
is determined by macroscopic hydrodynamic 
boundary conditions, notably the Hubble rate. 
In classical nucleation theory, the radius of a critical
bubble is $R \approx 2 \sigma / [L (1- T^{ }_\rmi{n}/\Tc^{ })]$, where $\sigma$
is the surface tension, $L$ is the latent heat, $\Tc^{ }$ is the critical
temperature, and $T^{ }_\rmi{n} < \Tc^{ }$ is the nucleation temperature. 
As the Hubble rate is suppressed by $T/\mpl^{ }$ compared with the 
temperature, where $\mpl^{ } \approx 1.22 \times 10^{19}_{ } \; \mbox{GeV}$
is the Planck mass, we proceed through the transition extremely slowly. 
Therefore, nucleations can happen close to $\Tc^{ }$, for instance 
at $T^{ }_\rmi{n} \sim (0.99 ... 0.999)\Tc^{ }$. This may lead to 
a scenario with a bubble wall width $\ell^{ }_\rmi{w} \sim 10/T$, 
bubble radius $R \sim 100/T$, 
and inter-bubble distance $\ell^{ }_{\B} \sim 10^{10}_{}/T$.
While the wall width is similar to the magnetic screening length, 
$\ell^{ }_\rmi{w}\sim \ell^{ }_\rmii{M}$, the radius $R$ could be
much larger. 

If we consider very large distances, 
$R \gg \ell^{ }_\rmii{M}$, 
the relevant degrees of freedom are the hydrodynamics ones. 
At these scales, gauge field fluctuations are exponentially screened. 
Only gauge-invariant quantities, associated with conserved currents like 
the energy-momentum tensor, can display correlated 
long-distance fluctuations. 

As a motivation for our investigation, we envisage a description
of phase transition kinetics in which an effective low-energy 
scalar field is added to the usual hydrodynamic variables (temperature, $T$, 
and flow velocity, $u^\mu_{ }$), to serve as an order parameter. The scalar
field is gauge neutral; we do not consider the full Higgs doublet
like in refs.~\cite{bbl1,bbl2}. That said, the neutral 
component of the Higgs field is essential for the dynamics, 
as it can become tachyonic around the transition. 
Therefore, we heuristically postulate that the neutral Higgs 
component 
serves as an additional hydrodynamic variable.\footnote{%
 This requires gauge fixing, however the coefficient that we  
 derive is shown to be gauge independent at the order of 
 the computation. 
 }  
We denote its value by $v$.

When $v$ is displaced from equilibrium, 
it relaxes towards it by dissipating its kinetic 
and potential energy into 
thermal entropy. Conversely, new fluctuations are constantly
created by kicks from the medium. This physics is captured
by a Langevin equation, 
\be
 \partial^2_{ } v - V'(v,T) 
 =  
       \frict^{ }_{}(v,T,... )
       \, u \cdot \partial v - \varrho
 \;, \la{langevin}
\ee
where the metric 
convention ($-$$+$$+$$+$) is assumed,\footnote{%
 We adopt this choice as it makes the analytic continuation
 between the imaginary-time formalism and the physical Minkowskian
 spacetime straightforward. 
 }  
$V$ is the Higgs potential, 
and $\varrho$ is a stochastic noise term, with the autocorrelator
\be
 \langle\, \varrho(\X) \varrho(\Y)\, \rangle = 
 \Omega( v,T,... ) \, \delta^{(4)}_{ }(\vec{\X-\Y})
 \;. \la{noise}
\ee 
The goal here is to estimate the value of the coefficient $\frict$.
This should simultaneously fix $\Omega$, 
through the fluctuation-dissipation
relation $\Omega \simeq 2 T \Upsilon$.

The dots in the arguments of $\frict$ and $\Omega$
in \eqs\nr{langevin} and \nr{noise}
stand for further gradients, such as 
$
 u\cdot\partial, \partial^2_{ }, ...\,
$.
The low-energy effective description is valid 
when the gradients are small.  
As this is the case for large bubbles, 
the Langevin description of \eq\nr{langevin}
can arguably be used to address nucleations
around the electroweak phase transition~\cite{lgv1,lgv2,hydro,ae}.  
It has also been postulated to provide for 
a description of the subsequent bubble growth~\cite{ikkl}
(though then the noise is normally omitted).
However, in the latter case, 
the gradients can be large within the bubble walls
($\partial \sim 1/\ell^{ }_\rmi{w}$), 
which draws the direct correspondence into question. 

\vspace*{3mm}

For a perturbative treatment, we write $v$ in \eq\nr{langevin}
as a sum of a background value ($\bar v$), 
and small fluctuations around it ($\delta v \equiv h$). 
We treat the background value as slowly varying, 
with $V'(\bar v,T) \simeq 0$. Then, \eq\nr{langevin} turns into
an equation for the fluctuations around the minimum. We denote
${m}_{h}^2 \equiv V''(\bar{v},T)$, remarking that this is only
one among several possible definitions of a thermally 
modified Higgs mass.\footnote{%
 This definition is not gauge independent at higher orders, 
 but sufficient for our leading-order computation. 
 } 

For the fluctuations, 
we go to momentum space, 
\be
 h(\X) 
 = \int_{\K} h(\K)\, e^{ i \K\cdot\X }_{ } 
 = \int_{\K} h(\K)\, e^{ - i \omega t + i \vec{k}\cdot\vec{x} }_{ }
 \;, \quad
 \K = (\omega,\vec{k})
 \;, 
\ee
and to a local rest frame, 
\be
 u^{\mu}_{ } \to (1,\vec{0})
 \;.
\ee
Then the Green's function (propagator)
corresponding to \eq\nr{langevin} satisfies
\be
 \bigl[ \omega^2 - k^2 - {m}_h^2 + i \omega \frict + ... \bigr] 
 \Delta^{-1}_{\K;h} = - 1
 \;, \quad 
 k \equiv |\vec{k}|
 \;, 
\ee
where the dots stand for higher powers of $\omega$ and $k$;
the overall sign is a convention, corresponding to 
a Euclidean propagator;
and the subscript $\K$ indicates that the frequency has been 
analytically continued to a Minkowskian one. 
Under the assumption 
\be 
 \frict \ll 
 \epsilon^{ }_h \; \equiv \; \sqrt{k^2 + {m}_h^2}
 \;, 
\ee
we then obtain $\Upsilon$ from the inverse propagator as 
\be
 \frict = -\, 
 \lim_{\vec{k}\to\vec{0}}
 \frac{\im \Delta^{ }_{\K;h}}{\omega}
 \;, \quad
 \omega \approx m^{ }_h
 \;, \la{Upsilon}
\ee
where the limit of small $\vec{k}$ should be understood in relation
to mass scales like $m^{ }_h$ and $gT$.

\section{Contribution of $2 \leftrightarrow 2$ processes}
\la{se:processes}

\subsection{Outline}
\la{ss:outline}

In vacuum, the dominant Higgs decay channels are 
$1\to 2$ processes into (almost) on-shell states, 
like $h\to b\bar{b}$ or $h\to\gamma\gamma$, or into 
virtual states that decay further, like 
$h\to W_{ }^{+*}W_{ }^{-*}$. At temperatures close to 
the electroweak transition, 
where 
according to~\eq\nr{regime}
the Higgs mass 
is smaller than the typical 
thermal masses, 
most of these channels close.\footnote{%
  Even the photons get a thermal mass $\sim eT$, and 
  the photon channel has a further loop suppression 
  factor, so that the corresponding rate 
  is arguably very small.
} 

It is possible, however, that the Higgs boson scatters off
plasma particles before its decay, setting it off-shell
(``brehmsstrahlung''). This turns the would-be $1\to 2$ decay
into a $1+n\to 2+n$ inelastic process, with $n\ge 1$. 
Alternatively,  we may pull one of 
the legs into the initial state, being then encountered
with a $2+n\to 1+n$ process. Simpler still, adding a single
thermal particle to a $1\to 2$ decay or 
$2\to 1$ inverse decay, yields a $2\to 2$ reaction
(or $1\to 3$ or $3 \to 1$, but these are phase-space constrained). 
It turns out that in an ultrarelativistic plasma, where all
masses are small compared with the temperature, the $2\to 2$ reactions
often represent the leading channel, given that they are 
always kinematically allowed. 
We consider them
in the present section, returning to the $1+n\leftrightarrow 2+n$
processes in \se\ref{se:1n_2n}.

Specifically, we discuss here {\em bosonic} $2\to 2$ reactions. 
The bosonic channels are important, because they can
be enhanced by the Bose distribution. They may
then develop an IR sensitivity to the Debye scale $\sim gT$, turning
the naive rate $\sim g^4_{ }T/\pi^3_{ }$
into 
$\sim (g^4_{ }T/\pi^3_{ }) \times (\pi T)^2_{ } / (g T)^2_{ }
\sim g^2_{ }T/\pi$.
The corresponding fermionic
channels, where the Bose enhancement is absent but
the strong gauge coupling makes an appearance, 
are discussed in appendix~A.
We start by considering the confinement phase ($T > \Tc^{ }$), where the Higgs
expectation value~$\bar v$ is set to zero (\se\ref{ss:confinement}), 
and then turn to the Higgs phase~(\se\ref{ss:higgs}). 

\subsection{Confinement phase}
\la{ss:confinement}

%
\begin{figure}[t]
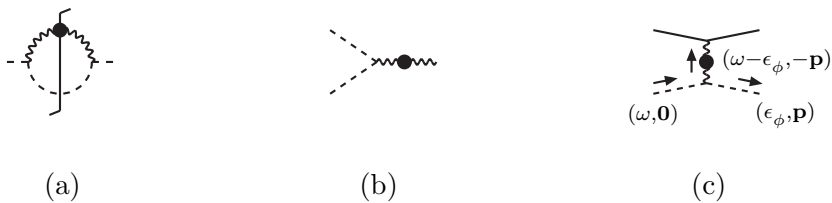


\begin{eqnarray}
&& 
 \hspace*{-0.5cm}
 \cutC \hspace*{25mm}
 \ampA \hspace*{25mm}
 \ampC
 \nn[8mm] 
&&            \hspace*{0.1cm} 
   \mbox{(a)} \hspace*{3.7cm}
   \mbox{(b)} \hspace*{3.9cm}
   \mbox{(c)}
 \nonumber
\end{eqnarray}

\vspace*{-3mm}

\caption[a]{\small 
 (a)~the imaginary part of a HTL-resummed Higgs self-energy
 contribution in the confinement (``symmetric'') phase. 
 Dashed lines denote scalar fields, wiggly lines gauge fields,
 and a blob HTL resummation.
 (b)~a~$2\to 1$ contribution originating from process~(a), 
 which is kinematically allowed in the regime of \eq\nr{regime}. 
 (c)~a~$2\to 2$ contribution originating from process~(a), 
 with the straight line standing
 for a generic particle species 
 (scalar, fermion, gauge field). 
 The kinematic variables correspond to those appearing 
 in \eq\nr{conf_1}, 
 with $\omega = m^{ }_\phi$ 
 and $\epsilon^{ }_\phi = \sqrt{p^2 + m_\phi^2}$. 
} 
\la{fig:symmetric}
\end{figure}
%

Anticipating that the momentum transfers pertinent
to $t$-channel $2\leftrightarrow 2$ scatterings will be soft
($p, \epsilon^{ }_\phi \ll \pi T$), we carry
out the computation with 
Hard Thermal Loop (HTL) resummed
perturbation theory~\cite{htl1,htl2,htl3,htl4}. 
The leading HTL
diagram contributing to the scalar field
self-energy at temperatures above the electroweak
phase transition ($T > \Tc^{ }$) 
is shown in \fig\ref{fig:symmetric}. 
In the confinement phase, as this regime is often referred to,  
we denote the scalar field by $\phi$ rather than $h$, 
and its (thermally corrected) 
mass by $m^{ }_\phi$ rather than $m^{ }_h$.
 We assume $m^{ }_\phi$ to be parametrically of the same order
 as $m^{ }_h$, as given by \eq\nr{regime}, 
 in order to in principle permit for 
 a comparison between the resulting widths in
 the two phases
 (cf.\ \se\ref{se:soft}).

The HTL-resummed gauge propagator 
appearing in the diagrams 
takes the form 
\be
 [\Delta^{-1}_{P;A}]^\rmii{ }_{\mu\nu} = 
 \frac{ [\PT_P]^{ }_{\mu\nu} }{P^2 + \Pi^\rmii{T}_{P} } +
 \frac{ [\PE_P]^{ }_{\mu\nu} }{P^2 
      + \Pi^\rmii{E}_{P}
    } +
 \frac{\xi P^{ }_\mu P^{ }_\nu}{P^4}
 \;, \la{htl_prop}
\ee
where $\xi$ is a gauge parameter; 
$P = (p^{ }_n,\vec{p})$ is an imaginary-time (Euclidean) 
four-momentum, 
with $p^{ }_n = 2\pi T n$
denoting a Matsubara frequency ($n\in\mathbbm{Z}$);
$A$ refers to the particle species (gauge field);
and the projectors read
\be
 [\PT_P]^{ }_{\mu\nu} 
 \; \equiv \; 
 \delta^{ }_{\mu i}\delta^{ }_{\nu j}
 \biggl(
  \delta^{ }_{ij} - \frac{p^{ }_i p^{ }_j}{p^2} 
 \biggr)
 \;, \quad
 [\PE_P]^{ }_{\mu\nu}
 \; \equiv \; 
 \delta^{ }_{\mu\nu} - 
 \frac{P^{ }_\mu P^{ }_\nu}{P^2_{ }} -
 [\PT_P]^{ }_{\mu\nu} 
 \;, \la{projectors}
\ee 
where we have denoted 
$
  p \; \equiv \; |\vec{p}|
$.
The self-energy $ \Pi^\rmii{E}_{P} $ 
in \eq\nr{htl_prop} turns out to 
contain an overall~$P^2_{ }$, and it is often convenient to 
factor it out, whereby we denote the dimensionless coefficient function as 
\be
 \widehat\Pi^\rmii{E}_{P} 
  \; \equiv \; 
 \frac{ \Pi^\rmii{E}_{P} }{ P^2_{ } }
 \;. 
\ee

In order to carry out the Matsubara sums appearing in the loop, 
the propagators can be written in a spectral representation, 
\be
 \frac{1}{P^2 + \Pi^\channel_{P}}
 = 
 \int_{-\infty}^{\infty} 
 \! 
 \frac{{\rm d}p^0_{ }}{\pi}
 \frac{\rho^\channel_{\P} }{p^0_{ }- i p^{ }_n}
 \;, \quad
 \channel \in \{ \rmi{T,E} \}
 \;, \la{spectral}
\ee
where $\P = (p^0_{ },\vec{p})$ is a Minkowskian four-momentum.
The spectral function is given by 
\be
 \rho^\channel_{\P} = 
 \im \biggl[ \frac{1}{P^2 + \Pi^\channel_{P} }
     \biggr]^{ }_{p^{ }_n \to -i (p^0_{ } + i 0^+_{ })}
 \;. \la{rho}
\ee
These representations hold if the Euclidean side is
a decreasing function of $p^{ }_n$, and the spectral function is 
a decreasing function of $p^0_{ }$.

After the analytic continuation in \eq\nr{rho}, 
the self-energies can be written as~\cite{qed1,qed2,qed3} 
\ba 
 \Pi^\rmii{T2}_{(-i(p^0_{ } + i 0^+_{ }),\vec{p})} & = & 
 \frac{\mE^2}{2} 
 \biggl\{ 
   \frac{(p^0_{ })^2}{{p}^2} + 
   \frac{p^0_{ }}{2p}
   \biggl[
     1 -  \frac{(p^0_{ })^2}{{p}^2}
   \biggr] 
   \ln\frac{p^0_{ }  + p + i 0^+_{ }}{p^0_{ } - p + i 0^+_{ }}
 \biggr\} 
 \;, \la{PiT} \\
 \widehat\Pi^\rmii{E2}_{(-i(p^0_{ } + i 0^+_{ }),\vec{p})} & = & 
 \frac{ \mE^2 }{{p}^2}
   \biggl[
     1 -     
     \frac{p^0_{ }}{2p}
   \ln\frac{p^0_{ } + p + i 0^+_{ }}{p^0_{ } - p + i 0^+_{ }}
   \biggr] 
 \;. \la{hatPiE}
\ea
Here the ``2'' stands
for SU$^{ }_\rmii{L}$(2) gauge bosons, 
and the thermal mass parametrizing
their HTL self-energies reads
\be
 \mE^2 = \biggl(  \fr23 + \fr{\nS}6 + \frac{\nG}{3} \biggr) g_2^2 T^2
 + \rmO(g_2^4)
 \;, 
\ee
where
$\nS^{ } \equiv 1$ is the number of Higgs doublets and 
$\nG^{ } \equiv 3$ the number of fermion generations.
The corresponding parameter for the 
U$^{ }_\rmii{Y}$(1) gauge bosons reads 
$m_\rmii{E1}^2
 = ( \fr{\nS}6 + \frac{5 \nG}{9} )g_1^2 T^2
$.
 
The self-energies need to be evaluated both 
in a spacelike domain (``$t$-channel'') and in a timelike
domain (``$s$-channel''). 
For $|p^0_{ }| < p$, 
both  self-energies have an imaginary part, 
originating from 
$
 \ln\frac{p^0_{ } + p + i 0^+_{ }}{p^0_{ } - p + i 0^+_{ }}
 \stackrel{\rmii{$ |p^0_{ }| < p $}}{=}
 - i \pi
$. 
In this situation we can write 
\ba 
 \Pi^\rmii{T2}_{(-i(p^0_{ } + i 0^+_{ }),\vec{p})} 
 & \stackrel{ }{=} & 
 \underbrace{ 
 \frac{\mE^2}{2} 
 \biggl\{ 
   \frac{(p^0_{ })^2}{{p}^2} + 
   \frac{p^0_{ }}{2p}
   \biggl[
     1 -  \frac{(p^0_{ })^2}{{p}^2}
   \biggr] 
   \ln\biggl| \frac{p^0_{ }  + p}{p^0_{ } - p} \biggr|
 \biggr\} 
 }_{ \equiv\; \Sigma^\rmii{T2}_{\P}}
 \nn 
 & - &
 \; i \,
 \underbrace{ 
 \frac{\pi \mE^2\, p^0_{ }}{4 p} 
   \biggl[
     1 -  \frac{(p^0_{ })^2}{{p}^2}
   \biggr] 
 }_{\equiv\; \Gamma^\rmii{T2}_{\P}}
  \, 
  \theta(p - |p^0_{ }|)
  \;, \la{PiT_split} \\[3mm]
 \widehat\Pi^\rmii{E2}_{(-i(p^0_{ } + i 0^+_{ }),\vec{p})} 
 & \stackrel{ }{=} & 
 \underbrace{ 
 \frac{ \mE^2 }{{p}^2}
   \biggl[
     1 -     
     \frac{p^0_{ }}{2p}
   \ln\biggl| \frac{p^0_{ } + p}{p^0_{ } - p } \biggr|
   \biggr] 
  }_{\equiv\; \widehat \Sigma^\rmii{E2}_{\P} }
 \; + \; 
 i \, 
 \underbrace{ 
 \frac{\pi \mE^2\, p^0_{ }}{2 p^3_{ }} 
 }_{\equiv\; \widehat \Gamma^\rmii{E2}_{\P}}
  \, 
 \theta(p - |p^0_{ }|)
 \;. \la{hatPiE_split}
\ea
In contrast, the self-energies are real
for $|p^0_{ }| > p$. 
Then the imaginary part for \eq\nr{rho} originates from the pole
induced by the tree-level frequency dependence, 
\be
 p_n^2 \;\to\; - (p^0_{ } + i 0^+_{ })^2 
       \;\supset\; - i \sign(p^0_{ }) 0^+_{ }
 \;,  
\ee 
after recalling that 
\be
 \im \biggl[ \, \frac{1}{
   f(x) - i g(x) 0^+_{ }  }
  \, \biggr] 
 \; = \;  
 \pi \sign\bigl(g(x)\bigr)\,\delta\bigl(f(x)\bigr)
 \; = \; \sum_\rmii{poles}
 \frac{ \pi \sign\bigl(g(x^{ }_\rmii{pole})\bigr)
 \,\delta\bigl(x-x^{ }_\rmii{pole}\bigr) }
 {|\, f'(x^{ }_\rmii{pole})\,|}
 \;, \la{pole}
\ee
where $x^{ }_\rmii{pole}$ satisfies $f(x^{ }_\rmii{pole})=0$.

The imaginary parts originating from 
\eqs\nr{PiT_split} and \nr{hatPiE_split} are frequently
referred to as ``cuts'', and those from \eq\nr{pole}
as ``poles''~\cite{dilepton}. 
As there are two lines in a loop
(cf.\ \fig\ref{fig:symmetric}), 
we may then find ``cut-cut'', ``pole-cut'', 
and ``pole-pole'' contributions. 
Such a general situation is encountered
in \se\ref{ss:higgs}, and 
will be illustrated
in \fig\ref{fig:kinematics}(left).

Returning to the diagram in \fig\ref{fig:symmetric}, 
inserting the propagators, 
and carrying out the contractions, 
we find the inverse propagator
\ba
 \Delta^{ }_{K;\phi}  
 & \supset &  
 K^2 + m_\phi^2 
 - \sum_{a=1}^{2} \frac{d^{ }_a g_a^2}{4}\,
   \Tint{P}
   \frac{(2 K^{ }_\mu - P^{ }_\mu)(2 K^{ }_\nu - P^{ }_\nu)}
        {(K-P)^2 + m_\phi^2}
 \la{Pi_phi_1} \\
 & \times & 
 \biggl\{ 
 \delta^{ }_{\mu i} \delta^{ }_{\nu j}
 \biggl(
  \delta^{ }_{ij} - \frac{p^{ }_i p^{ }_j}{p^2} 
 \biggr)
 \biggl[
   \frac{1}{P^2 + \Pi^\rmii{T$a$}_{P}} 
  - 
   \frac{1}{P^2 + \Pi^\rmii{E$a$}_{P}} 
 \biggr]
 + 
 \biggl(
  \frac{\delta^{ }_{\mu\nu}}{P^2} - \frac{P^{ }_\mu P^{ }_\nu}{P^4} 
 \biggr)
 \frac{1}{1 + \widehat \Pi^\rmii{E$a$}_{P}}
 \biggr\}
 \;, \nonumber  
\ea
where 
$
 d^{ }_1 \equiv 1
$
and
$ 
d^{ }_2 \equiv 3
$.
It is important to note that 
the part of \eq\nr{htl_prop}
containing the gauge parameter has dropped out. The reason for this
is that contributions originating from $P^{ }_\mu P^{ }_\nu$ 
can be written as 
\ba
 \frac{(P^2 - 2 K\cdot P)^2}{ (K-P)^2 + m_\phi^2 }
 & = & 
 \frac{(P^2 - 2 K\cdot P)[ (K-P)^2 + m_\phi^2 - (K^2 + m_\phi^2) ]}
 {(K-P)^2 + m_\phi^2}
 \nn 
 & = & 
 \underbrace{P^2}_{\rm no~\it K} 
 \; - \; \underbrace{2 K\cdot P}_{\rm odd~in~\it P}
 \; - \; \underbrace{(K^2 + m_\phi^2)}_{\rm vanishes~on-shell} 
   \frac{P^2 - 2 K\cdot P}{(K-P)^2 + m_\phi^2}
 \;. \la{gauge_part}
\ea
Here the first term is non-zero, but since it is independent of $K$, 
it has no imaginary part, and therefore gives no contribution 
to \eq\nr{Upsilon}. The second term drops out because of its 
antisymmetry, the third because the scalar 
self-energy is evaluated on-shell.

In the first term on 
the second line of \eq\nr{Pi_phi_1}, the projector eliminates the
dependence on $P^{ }_\mu$ and $P^{ }_\nu$. From $K^{ }_\mu$ and $K^{ }_\nu$, 
only the spatial part is left over. But as we are interested in the limit
$\vec{k}\to \vec{0}$ (cf.\ \eq\nr{Upsilon}), these drop out as well. 

In the second term on 
the second line of \eq\nr{Pi_phi_1}, 
which originates exclusively from $\delta^{ }_{\mu\nu}/P^2$
due to the argument in \eq\nr{gauge_part},
we may write 
\ba
 \frac{(2 K - P)^2}{ (K-P)^2 + m_\phi^2 }
 & = & 
 \frac{4 K^2 + P^2 + 2 [ (K-P)^2 + m_\phi^2 - P^2 - (K^2 + m_\phi^2) ]}
 {(K-P)^2 + m_\phi^2}
 \nn 
 & = & 
 \underbrace{2}_{\rm no~\it K} 
 \; - \; \underbrace{(K^2 + m_\phi^2)}_{\rm vanishes~on-shell} 
   \frac{ 2 }{(K-P)^2 + m_\phi^2}
 \; + \; 
 \frac{4K^2 - P^2}{ (K-P)^2 + m_\phi^2 }
 \;. \hspace*{6mm}
\ea
Setting $K^2 \to -m_\phi^2$, we thus find
\ba
 \Delta^{ }_{(-i m^{ }_\phi,\vec{0});\phi} 
 & \supset &  
  \sum_{a=1}^2
  \frac{d_a^{ } g_a^2}{4}\,
   \Tint{P}
   \biggl(  \frac{4 m_\phi^2}{P^2} + 1 \biggr) 
   \frac{1}{1 + \widehat \Pi^\rmii{E$a$}_{P}}
   \frac{1}
        {(K-P)^2 + m_\phi^2}
   \bigg|^{ }_{K = (  -i m^{ }_\phi,\vec{0}  )}
 \;. \hspace*{5mm} \la{symm_1}
\ea

In order to carry out the Matsubara sum
in \eq\nr{symm_1}, we insert \eq\nr{spectral}, 
and denote 
\be
 \epsilon^{ }_\phi
  \;\equiv\; 
 \sqrt{p^2 + m_\phi^2}
 \;. \la{e_phi}
\ee
For later reference, we treat the Higgs propagator as if it had 
a general self-energy. 
Then
\ba
 && \hspace*{-1.0cm} 
 T\sum_{p^{ }_n}
 \frac{1}{
          [ P^2_{ } + \Pi^\channel_{ P } ] 
          [ (k^{ }_n - p^{ }_n)^2 + \epsilon_\phi^2 + \Pi^\phi_{ K-P } ]
         }
 \nn 
 & \stackrel{\beta = 1/T}{=} &
 T\sum_{p^{ }_n} T\sum_{q^{ }_n}  
 \beta \delta^{ }_{0,p^{ }_n + q^{ }_n - k^{ }_n}
 \frac{1}{ 
           [ p_n^2 + p^2 + \Pi^\channel_{ P } ] 
           [ q_n^2 + \epsilon_\phi^2 + \Pi^\phi_{ (q^{ }_n,\vec{k-p}) } ]
         }
 \nn 
 & \stackrel{\rmii{\nr{spectral}}}{=} &
 T\sum_{p^{ }_n} T\sum_{q^{ }_n}  
 \int_0^{\beta} \! {\rm d}\tau \, 
 e^{i(p^{ }_n + q^{ }_n - k^{ }_n)\tau}_{ } \, 
 \int_{-\infty}^{\infty}\! \frac{{\rm d}p^0_{ }}{\pi} 
 \int_{-\infty}^{\infty}\! \frac{{\rm d}q^0_{ }}{\pi} 
 \frac{ \rho^\channel_{ (p^0_{ },\,\vec{p}) } }{p^0_{ } - i p^{ }_n } \,
 \frac{ \rho^{\phi}_{ (q^0_{ },\,\vec{k-p}) }}{q^0_{ } - i q^{ }_n } 
 \nn 
 & = & 
 \int_{-\infty}^{\infty}\! \frac{{\rm d}p^0_{ }}{\pi} 
 \int_{-\infty}^{\infty}\! \frac{{\rm d}q^0_{ }}{\pi} 
 \, 
 \rho^\channel_{ (p^0_{ },\,\vec{p}) } \,
 \rho^{\phi}_{ (q^0_{ },\,\vec{k-p}) } \, 
 \int_0^{\beta} \! {\rm d}\tau \, 
 e^{- ik^{ }_n\tau}_{ } \, 
 \nB^{ }(p^0_{ }) e^{\tau p^0_{ }}_{ }
 \nB^{ }(q^0_{ }) e^{\tau q^0_{ }}_{ }
 \nn[2mm] 
 & = & 
 \int_{-\infty}^{\infty}\! \frac{{\rm d}p^0_{ }}{\pi} 
 \int_{-\infty}^{\infty}\! \frac{{\rm d}q^0_{ }}{\pi} 
 \, 
 \rho^\channel_{ (p^0_{ },\,\vec{p}) } \,
 \rho^{\phi}_{ (q^0_{ },\,\vec{k-p}) } \, 
 \frac{1 + 
 \nB^{ }(p^0_{ }) + 
 \nB^{ }(q^0_{ }) 
 }{p^0_{ } + q^0_{ } - i k^{ }_n}
 \;, 
\ea
where 
$
 \nB^{ }(x) \; \equiv \; 1/(e^{x/T} - 1)
$
is the Bose distribution. 
Subsequently we analytically continue 
$
 k^{ }_n \to -i [\omega + i 0^+_{ }]
$, 
and take the imaginary part,
\ba
 && \hspace*{-1.5cm} 
 \im \biggl\{ 
 T\sum_{p^{ }_n}
 \frac{1}{
          [ P^2_{ } + \Pi^\channel_{ P } ] 
          [ (k^{ }_n - p^{ }_n)^2 + \epsilon_\phi^2 + \Pi^\phi_{ K-P } ]
         }
 \biggr\}^{ }_{  k^{ }_n \to -i [\omega + i 0^+_{ }] }
 \nn 
 & = & 
 \int_{-\infty}^{\infty}\! \frac{{\rm d}p^0_{ }}{\pi} 
 \, 
 \rho^\channel_{ (p^0_{ },\,\vec{p}) } \,
 \rho^{\phi}_{ (\omega - p^0_{ },\,\vec{k-p}) } \, 
 \bigl[\, 
 1 + 
 \nB^{ }(p^0_{ }) + 
 \nB^{ }(\omega - p^0_{ }) 
 \,\bigr]
 \;. \la{matsubara_sum}
\ea
For $\vec{k} = \vec{0}$, 
the physical value is $\omega = m^{ }_\phi$
(cf.\ \eq\nr{symm_1}), but we keep 
the notation $\omega$ for the time being, because later on we 
apply the same formula for $\omega = m^{ }_h$. We note
that the derivation also applies to $1/(1 + \widehat\Pi^\alpha_P)$,
as long as this is a decreasing function of $p^{ }_n$, yielding
then a corresponding spectral function 
$\widehat\rho^{\hspace*{0.6mm}\alpha}_\P$.

Returning to the specific case of \eq\nr{symm_1}, where 
$\Pi^\phi_{K-P} = 0$, we make use of the free spectral function 
\be
 \rho^{\phi}_{(q^0_{ },\vec{k-p})}
 \underset{\vec{k}=\vec{0}}{\overset{\rmii{\nr{e_phi}}}{=}} 
 \frac{\pi}{2\epsilon^{ }_\phi}
 \bigl[\,
  \delta(q^0_{ } - \epsilon^{ }_\phi) - 
  \delta(q^0_{ } + \epsilon^{ }_\phi) 
 \,\bigr]
 \;. \la{rho_phi}
\ee
Writing
$
 \nB^{ }(-\epsilon^{ }_\phi) 
 = - 1 - \nB^{ }(\epsilon^{ }_\phi)
$, 
the result can then be expressed as
\ba
 && \hspace*{-1.5cm}
 \im\biggl[ 
 T\sum_{p^{ }_n}
  \frac{1}{P^2 + \Pi^\channel_{P}}
  \frac{1}{(K-P)^2 + m_\phi^2}
 \biggr]^{ }_{ K \to (-i [\omega + i 0^+_{ }],\vec{0}) }
 \la{conf_general} \\[2mm]
 & = &   
 \int_{-\infty}^{\infty}\! {\rm d}p^0_{ }\, 
 \frac{\rho^\channel_{\P}}{2\epsilon^{ }_\phi}
 \,\biggl\{
   \bigl[ 1 + \nB^{ }(p^0_{ })  + \nB^{ }(\epsilon^{ }_\phi) \bigr] 
   \,
   \delta(\omega - \epsilon^{ }_\phi - p^0_{ })
   \hspace*{2.0cm} \ampTmini
 \nn 
 & & \hspace*{2.0cm} \; + \, 
   \bigl[ \nB^{ }(\epsilon^{ }_\phi) - \nB^{ }(p^0_{ })\bigr]
   \, 
   \delta(\omega + \epsilon^{ }_\phi - p^0_{ })
 \biggr\}
 \;,
 \hspace*{2.1cm} \ampSmini
 \nonumber
\ea
with the graphs illustrating typical processes
for the terms on each corresponding line. 
The wiggly line, connecting to the grey blob, 
denotes an HTL resummed spectral function $\rho^\channel_\P$.   
The first case corresponds to $t$-channel exchange ($|p^0_{ }| < p$), 
the second to an $s$-channel reaction ($p^0_{ } > p$).
The $s$-channel is realized if a ``plasmon'' goes
on-shell; we return to this 
in \se\ref{se:confinement_1n_2n}.

Restricting to the $t$-channel part for now, 
\eq\nr{symm_1} becomes\hspace*{0.3mm}\footnote{%
 We remark that in the domain of validity of the HTL theory, 
 i.e.\ $\epsilon \ll \pi T$, we can expand 
 $ \nB^{ }(\epsilon) \approx \tfr{T}{\epsilon} - \tfr{1}{2}$, 
 however we keep the general notation for the time being, 
 to display the physical origin of the various terms. 
 The proper limiting value is adopted in the numerical 
 evaluation, cf.\ \eq\nr{soft_limit_conf}.   
 }
\ba
 && \hspace*{-0.5cm} 
 \im\Delta^{ }_{(-i [m^{ }_\phi + i 0^+_{ }],\vec{0});\phi}
 \; \supset \; 
 - \sum_{a=1}^{2} \frac{d^{ }_a g_a^2}{4} 
  \la{conf_1} \\
 & & \; \times \, 
 \int_\vec{p} 
 \frac{ 
   1 + \nB^{ }(\epsilon^{ }_\phi)
     + \nB^{ }(m^{ }_\phi - \epsilon^{ }_\phi)  
      }
                            {2\epsilon^{ }_\phi}
 \biggl[\frac{4m_\phi^2}
 {p^2 - (m^{ }_\phi - \epsilon^{ }_\phi)^2} + 1 \biggr]
 \frac{ 
 \widehat\Gamma^\rmii{E$a$}_{( m^{ }_\phi - \epsilon^{ }_\phi ,p)} }
 {
 [ 1 + \widehat\Sigma^\rmii{E$a$}_
   {( m^{ }_\phi - \epsilon^{ }_\phi ,p)} ]^2_{ }
 + 
 [ \widehat\Gamma^\rmii{E$a$}_{( m^{ }_\phi - \epsilon^{ }_\phi ,p)} ]^2_{ }
 }
 \;. \nonumber
\ea
The corresponding kinematics is 
illustrated in \fig\ref{fig:symmetric}(right).
We note that 
$
 |m^{ }_\phi - \sqrt{m_\phi^2 + p^2_{ }}| < p
$, 
so the latter factor is indeed in a spacelike domain. 
A limiting value of this finite integral 
will be evaluated in \se\ref{ss:soft_2to2}. 

\subsection{Higgs phase}
\la{ss:higgs}

%
\begin{figure}[t]
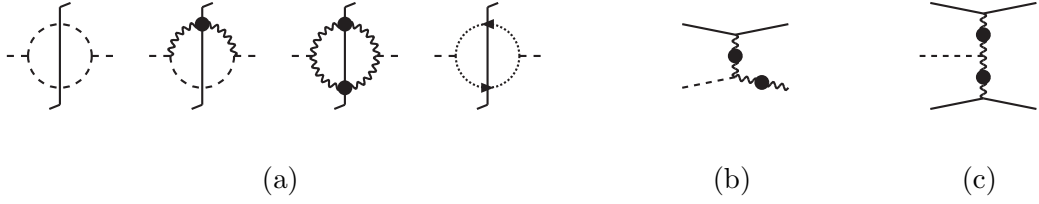


\begin{eqnarray}
&& 
 \hspace*{-1.5cm}
 \cutHa \hspace*{1mm}
 \cutHb \hspace*{1mm}
 \cutHc \hspace*{1mm}
 \cutHd \hspace*{15mm}
 \ampBa  \hspace*{15mm}
 \ampBBa
 \nn[8mm] 
&&            \hspace*{2.0cm} 
   \mbox{(a)} \hspace*{5.5cm}
   \mbox{(b)} \hspace*{2.8cm}
   \mbox{(c)}
 \nonumber
\end{eqnarray}

\vspace*{-4mm}

\caption[a]{\small 
 (a)~imaginary parts of a HTL-resummed Higgs self-energy 
 in the Higgs (``broken'') phase. 
 Dashed lines denote scalar fields, wiggly lines gauge fields,
 dotted lines ghosts, and a blob HTL resummation. 
 (b)~$2\leftrightarrow 2$ reaction originating 
 from the processes in~(a), with the solid line representing
 any particle species.
 (c)~$3\leftrightarrow 2$ reaction originating 
 from the processes in~(a).
} 
\la{fig:broken}
\end{figure}
%

We proceed to compute the Higgs width 
in the phase with a non-zero expectation value, $\bar v > 0$. 
The corresponding diagrams are shown in \fig\ref{fig:broken}. 
The computation is technically more cumbersome than in the 
confinement phase, but it yields a non-trivial demonstration of 
the gauge independence of $\Upsilon$, as well as of its partial
continuity across the phase transition.  

The HTL propagator of \eq\nr{htl_prop} gets modified in the 
presence of electroweak symmetry breaking. If $g\bar v \sim \mE^{ }$, 
i.e.\ $\bar v\sim T$, vacuum masses 
are of the same order as  
thermal corrections from HTL self-energies.
Then the propagator of a $W$-boson 
in a general $R^{ }_\xi$ gauge can be written as  
\ba
 [\Delta^{-1}_{P;W}]^\rmii{ }_{\mu\nu}
 & = &  
 \underbrace{ 
 \frac{ \PT_{\mu\nu} }{P^2 + \Pi^\rmii{T2}_{P}  + \mW^2 } 
 + 
 \biggl[ 
   \delta^{ }_{\mu\nu} 
 + \frac{P^{ }_\mu P^{ }_\nu ( 1 + \widehat \Pi^\rmii{E2}_{P} )}{\mW^2}
 - \PT_{\mu\nu} 
 \biggr]
 \frac{1}{P^2 (1 + \widehat \Pi^\rmii{E2}_{P}) + \mW^2}
 }_{\equiv~ \rm physical~part}
 \nn 
 & - & 
 \underbrace{  
 \frac{P^{ }_\mu P^{ }_\nu}{\mW^2 (P^2 + \xi \mW^2)}
 }_{\equiv~ \rm gauge~part}
 \;. \la{htl_prop_w}
\ea
The presentation is chosen so that there is only one pole
in each part (no $1/P^2_{ }$).
The gauge boson vacuum mass is given by $\mW^{ } \equiv g^{ }_2\bar{v}/2$, 
and similarly $\mZ^{ } \equiv \sqrt{g_1^2 + g_2^2}\,\bar{v}/2$.

The gauge propagators are more complicated in the neutral 
sector ($\gamma,Z^0_{ }$).
The reason is that the thermal self-energies do not ``align'' with
the way that vacuum masses appear, so that the diagonalization of
the $2\times 2$ propagator matrix, and correspondingly 
the mixing angles, get modified in a $p$-dependent way
(cf.\ appendix~B of ref.~\cite{broken}). 

More precisely, 
\eq\nr{htl_prop_w} contains three parts: 
the {\sc T}-part, 
the {\sc E}-part, 
and the gauge part. 
The mixing angles change in different ways 
in the {\sc T} and {\sc E}-parts, because 
$\Pi^\rmii{T\hspace*{-0.3mm}$a$}_{ }$ and 
$\widehat \Pi^\rmii{E$a$}_{ }$ are different
($a=1,2$). Moreover, 
the mixing angles are momentum-dependent, because 
$\Pi^\rmii{T\hspace*{-0.3mm}$a$}_{ }$ and 
$\widehat \Pi^\rmii{E$a$}_{ }$ are so. The mixing angles do 
{\em not} get modified in the gauge part, because no thermal
self-energy appears there. 
Similarly, the ghost and Goldstone
sectors have no momentum-dependent self-energies. 
In the following, we do not display the contributions
of the neutral sector explicitly, but just refer to 
them collectively as the ``$Z$-channel''.

Proceeding with the computation, the first step is to check 
the cancellation of gauge dependence. Gauge-dependent pole locations,
such as $\xi\mW^2$ in \eq\nr{htl_prop_w}, originate from the
vector, ghost, and Goldstone propagators. Adding all terms
and going on-shell ($K^2 = -m_h^2 = -2 \lambda \bar v^2_{ }$), all 
momentum-dependent terms cancel. Therefore the remainder has 
no imaginary part in the sense of \eq\nr{Upsilon}
(the momentum-independent gauge terms would be cancelled by
further Feynman diagrams, not shown in \fig\ref{fig:broken}). 

Turning to the gauge independent terms, 
and taking already the limit $k\to 0$, we find 
\ba
 \Delta^{ }_{(-i m^{ }_h,\vec{0});h} 
 & \supset &  
   - g_2^2 \,
   \Tint{P}
   \biggr\{ 
   \frac{2 \mW^2}{[P^2 + \Pi^\rmiii{T2}_{P} + \mW^2]
                  [(K-P)^2 + \Pi^\rmiii{T2}_{K-P} + \mW^2]}
 \nn
 & + & 
   \frac{1}{\mW^2}
   \frac{(P^2 - K\cdot P)^2
        (1 + \widehat\Pi^\rmiii{E2}_{P})
        (1 + \widehat\Pi^\rmiii{E2}_{K-P}) }
   {
   [ P^2 ( 1 + \widehat\Pi^\rmiii{E2}_{P} ) + \mW^2 ] 
   [ (K-P)^2 ( 1 + \widehat\Pi^\rmiii{E2}_{K-P} ) + \mW^2 ] 
   }
 \biggr\}^{ }_{K = (  -i m^{ }_h,\vec{0}  )}
 \nn[2mm]
 & + & \mbox{($Z$-channel)}
 \;. \hspace*{5mm} \la{broken_2}
\ea
To identify the origin of the second line of \eq\nr{broken_2}
in terms of the structures in \eq\nr{htl_prop_w}, 
we remark that the vertex $\sim h W^+_{ }W^-_{ }$ gives a factor
$\sim \mW^2$, and the {\sc E}-part
has a prefactor $\sim 1/\mW^4$, 
from two appearances on the first line 
of \eq\nr{htl_prop_w}, 
which combine to give $\sim 1/\mW^2$.

%
\begin{figure}[t]

\centerline{
  \epsfysize=9.0cm\epsfbox{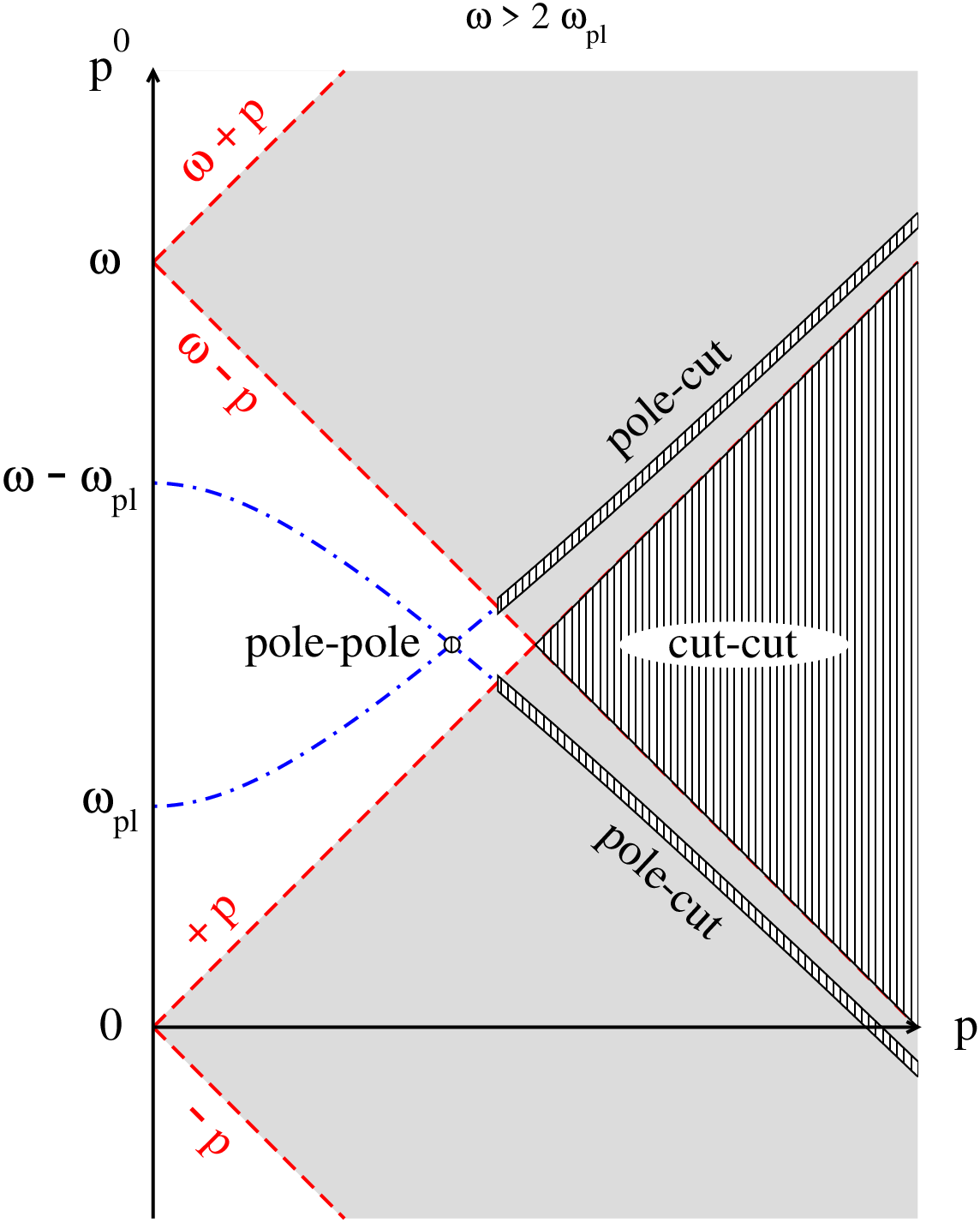}
  \hspace*{4mm}
  \epsfysize=9.0cm\epsfbox{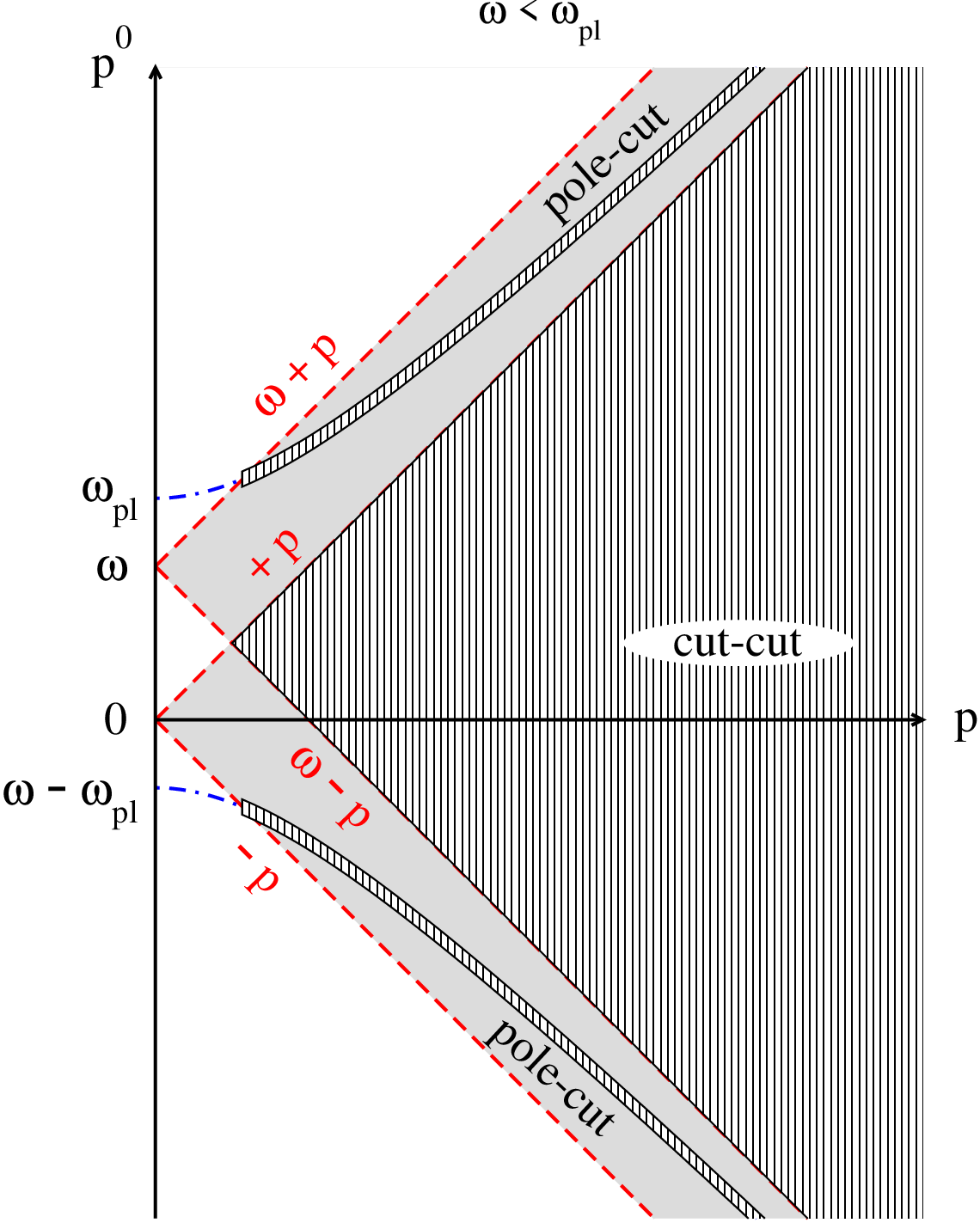}
}

\vspace*{1mm}

\caption[a]{\small 
 The kinematics pertinent to \eq\nr{higgs_2}, 
 with $\omega^{ }_\rmii{pl}$ denoting a ``plasmon mass'' 
 (cf.\ the text). 
 In the left panel, 
 we show a situation with 
 $\omega = m^{ }_h > 2 \omega^{ }_\rmii{pl}$, so that a pole-pole 
 contribution exists. 
 In the right panel, 
 we assume $\omega = m^{ }_h < \omega^{ }_\rmii{pl}$, 
 cf.\ \eq\nr{regime}, 
 and then it is absent.
 The integration domain of \eq\nr{higgs_2} selects the lower 
 pole-cut branch. The integration over the cut-cut domain
 can likewise be symmetrized, 
 cf.\ \eq\nr{higgs_3}. 
} 
\la{fig:kinematics}
\end{figure}
%

The Matsubara sum in \eq\nr{broken_2}
can be carried out with the help 
of \eq\nr{matsubara_sum}, 
once we insert \eq\nr{spectral}.\footnote{%
 As Matsubara frequencies now appear in the numerator, 
 we are faced also with structures of the type 
 $ 
 { p_n^2 } / [{p_n^2 + p^2 + \Pi^\channel_P}]
 $. 
 Given that this does not decrease at large $|p^{ }_n|$, 
 the spectral representation from 
 \eqs\nr{spectral} and \nr{rho}
 is not literally applicable. 
 However, we can add and subtract $p^2 + \Pi^\channel_P$ in the numerator, 
 so that the problematic part cancels against the denominator. 
 The resulting ``large'' 1 yields no cut. 
 By carrying out a tedious analysis of the other terms,
 it can be verified that the correct result can in fact
 be obtained 
 with the naive replacement
 $p^{ }_n \to - i p^0_{ }$.
 In the course of the proof, we need to note that
 in the domain $|p^0_{ } | < p$, 
 \ba
 \im^{ }_{\rmiii{(cut)}}\biggl[  
   \frac{ 1 + \widehat\Pi^\rmiii{E2}_{P} }
   {
    P^2 ( 1 + \widehat\Pi^\rmiii{E2}_{P} ) + \mW^2  
   }\biggr]^{ }_{p^{ }_n \to -i [p^0_{ } + i 0^+_{ }]}
 & = & 
 \im\biggl[  
   \frac{ 1 + \widehat\Sigma^\rmiii{E2}_{\P}
  + i \widehat\Gamma^\rmiii{E2}_{\P} }
   {
    \P^2_{ } 
        ( 1 + \widehat\Sigma^\rmiii{E2}_{\P}
  + i \widehat\Gamma^\rmiii{E2}_{\P}
        ) + \mW^2  
      }\biggr]
 \nn
   = \; 
 \im\biggl\{\,  
   \frac{ 
   [ 1 + \widehat\Sigma^\rmiii{E2}_{\P}
    + i \widehat\Gamma^\rmiii{E2}_{\P} ]
   [ 
    \P^2_{ } ( 1 + \widehat\Sigma^\rmiii{E2}_{\P} 
        ) + \mW^2  
    - i \P^2_{ }\widehat\Gamma^\rmiii{E2}_{\P}
   ]
   }
   {
   [ \P^2_{ } ( 1 + \widehat\Sigma^\rmiii{E2}_{\P} 
        ) + \mW^2 ]^2_{ } + 
   ( \P^2_{ } \widehat\Gamma^\rmiii{E2}_{\P} )^2_{ } 
   }
 \,\biggr\}
 & = & 
   \frac{ 
     \mW^2 \widehat\Gamma^\rmiii{E2}_{\P}
   }
   {
   [ \P^2_{ } ( 1 + \widehat\Sigma^\rmiii{E2}_{\P} 
        ) + \mW^2 ]^2_{ } + 
   ( \P^2_{ } \widehat\Gamma^\rmiii{E2}_{\P} )^2_{ } 
   }
 \;. 
 \hspace*{5mm}
       \la{cut_PiE}
 \ea 
 Incidentally, \eq\nr{cut_PiE} shows that 
 \be
 \im^{ }_{\rmiii{(cut)}}\biggl[  
   \frac{ 1 + \widehat\Pi^\rmiii{E2}_{P} }
   {
    P^2 ( 1 + \widehat\Pi^\rmiii{E2}_{P} ) + \mW^2  
   }\biggr] 
 = 
 - \frac{\mW^2}{\P^2_{ }}
 \im^{ }_{\rmiii{(cut)}}\biggl[  
   \frac{ 1 }
   {
    P^2 ( 1 + \widehat\Pi^\rmiii{E2}_{P} ) + \mW^2  
   }\biggr] 
 \;. \la{rel_1}
 \ee
 Remarkably, \eq\nr{rel_1} holds also for pole 
 contributions, 
 because in this case 
 $ P^2 ( 1 + \widehat\Pi^\rmiii{E2}_{P} ) + \mW^2 = 0$. 
 }
In \eq\nr{matsubara_sum}, each propagator is represented
by a corresponding spectral function.  
The spectral functions are non-vanishing for 
$|p^0_{ }| < p$ and $|\omega - p^0_{ }| < p$ (cut parts),  
as well as 
$|p^0_{ }| > p$ and $|\omega - p^0_{ }| > p$ (pole parts), 
respectively.
In this section we focus on the pole-cut contributions
(the kinematics is illustrated in \fig\ref{fig:kinematics}). 

Given that two same spectral functions appear, 
we can symmetrize the contributions, by picking up the cut parts
from $\rho^\channel_\P$ and the pole parts from $\rho^\channel_{\K-\P}$.
This fixes the integration domain of $p^0_{ }$, yielding
\ba
 && \hspace*{-1.0cm} 
 \im\Delta^{ }_{(-i [\omega + i 0^+_{ }],\vec{0});h}
 \; \underset{\K \, \equiv \, (\omega + i 0^+_{ },\vec{0})}
             {\overset{\P^2 \,\equiv\, p^2 - (p^0_{ })^2}{\supset}}
 \; 
 - \, 2 g_2^2 
 \int_\vec{p}
 \int_{-p}^{\min(p,\omega-p)}
 \! \frac{{\rm d}p^0_{ } }{\pi}
 \bigl[\,
  1 + \nB^{ }(p^0_{ }) + \nB^{ }(\omega - p^0_{ })
 \,\bigr]
 \nn[3mm] 
 & \times & \biggl\{ \, 
 \frac{2 \mW^2 \Gamma^\rmiii{T2}_{\P} 
      }
      { (\P^2 + \Sigma^\rmiii{T2}_{\P} + \mW^2)^2_{ } 
      + (\Gamma^\rmiii{T2}_{\P})^2}
 \im^{ }_\rmii{(pole)} \biggl[\,
     \frac{1}{ (\K-\P)^2_{ }
      + \Sigma^\rmiii{T2}_{\K-\P} 
      + \mW^2 } 
     \biggr] 
 \nn
 & + & 
 \underbrace{  
 \frac{[p^2 + p^0_{ }(\omega - p^0_{ })]^2_{ } \widehat \Gamma^\rmiii{E2}_{\P} 
      }
      { [\P^2(1 + \widehat\Sigma^\rmiii{E2}_{\P}) 
      + \mW^2]^2_{ } 
      + (\P^2 \widehat \Gamma^\rmiii{E2}_{\P})^2
      }
 }_{
  \rm cut~part
 }
 \underbrace{
 \im^{ }_\rmii{(pole)} \biggl[\,
     \frac{1
          + \widehat\Sigma^\rmiii{E2}_{\K-\P} 
          }
          {
         (\K-\P)^2_{ } 
         ( 1 
           + \widehat\Sigma^\rmiii{E2}_{\K-\P} 
         )
             + \mW^2 }  
     \biggr] 
  }_{
   \rm pole~part
  }
  \,\biggr\}
 \nn[2mm]
 & + & \mbox{($Z$-channel)}
 \;. \hspace*{5mm} \la{higgs_2}
\ea

Now, unlike in the confinement phase (cf.\ \eq\nr{rho_phi}), the pole part
cannot be given in closed form. 
However, 
making use of \eqs\nr{PiT_split} and \nr{hatPiE_split}, 
and denoting by $p^0_\alpha(p) > 0 $ the location of a pole, 
with $\alpha \in \{ \rmi{T,E} \} $, 
\eq\nr{pole} allows us to write
\ba
 \im^{ }_\rmii{(pole)} \biggl[\,
     \frac{1 
          }
          {
         \P^2_{ } 
           +  \Sigma^\rmiii{T2}_{\P} 
             + \mW^2 }  
     \biggr] 
 &
    \underset{ \P \,=\, (p^0_{ } + i 0^+_{ },\vec{p}) }
             {\overset{\P^2 \,\equiv\, p^2 - (p^0_{ })^2}{ \supset }}
 & 
     \sum_{\pm}
     \frac{
          \pi \sign(p^0_{ }) \delta(p^0_{ } \pm p^0_\rmiii{T2})
          }
          { 
         | 
         \partial^{ }_{p^0_{ }} \{ 
         \P^2_{ } 
           + \Sigma^\rmiii{T2}_{\P} 
          + \mW^2 
         \} 
         | 
         }
  \nn[3mm] 
  & \stackrel{ p^0_{ } > 0 }{\supset} & 
  \frac{
       \pi \delta(p^0_{ } - p^0_\rmiii{T2})
       \, p^0_\rmiii{T2} \, |\P^2_{ }|
       }{
       |\,
        \mW^2 [3 (p^0_\rmiii{T2})^2 - p^2_{ } ] 
       - \P^4_{ }
       + m^2_\rmiii{E2} (p^0_\rmiii{T2})^2_{ }
       \,|
       }
      \;, \la{jacob_T} \hspace*{6mm}
 \\[3mm]
 \im^{ }_\rmii{(pole)} \biggl[\,
     \frac{1 + \widehat\Sigma^\rmiii{E2}_{\P} 
          }
          {
         \P^2_{ } 
         ( 1 
           + \widehat\Sigma^\rmiii{E2}_{\P} 
         )
             + \mW^2 }  
     \biggr] 
 &
    \underset{ \P \,=\, (p^0_{ } + i 0^+_{ },\vec{p}) }
             {\overset{\P^2 \,\equiv\, p^2 - (p^0_{ })^2}{ \supset }}
 & 
     \sum_{\pm}
     \frac{
          \pi \sign(p^0_{ }) \delta(p^0_{ } \pm p^0_\rmiii{E2})
          \, ( 1 + \widehat\Sigma^\rmiii{E2}_{\P} )
          }
          { 
         |\, 
         \partial^{ }_{p^0_{ }} \{ 
         \P^2_{ } 
         ( 1 
           + \widehat\Sigma^\rmiii{E2}_{\P} 
         ) + \mW^2 
         \} 
         \,| 
         }
  \nn[3mm] 
  & \stackrel{ p^0_{ } > 0 }{\supset} & 
  \frac{
       \pi \delta(p^0_{ } - p^0_\rmiii{E2})
       \, p^0_\rmiii{E2} \, \mW^2 
       }{
       |\,
        \mW^2 [3 (p^0_\rmiii{E2})^2 - p^2_{ } ] 
       - \P^4_{ } 
       -  m^2_\rmiii{E2} \P^2_{ }  
       |\,
       }
      \;. \la{jacob_E} \hspace*{6mm}
\ea

To find the values of $p^0_\alpha$, which are functions of $p$, 
we need to determine numerically the zeros of the denominators, 
\ba
 p^2_{ } - (p^{0}_\rmii{T2})^2_{ } + 
 \Sigma^\rmii{T2}_{(p^{0}_\rmiii{T2},p)}
 + \mW^2 & = & 0
 \;, \la{zero_T2}
 \\[2mm]
 \bigl[
   p^2_{ } - (p^{0}_\rmii{E2} )^2 
 \bigr]
 \bigl[
  1 +  
 \widehat\Sigma^\rmii{E2}_{(p^{0}_\rmiii{E2},p)}
 \bigr]
 + \mW^2 & = & 0
 \;.  \la{zero_E2}
\ea
In \eq\nr{zero_E2}, the existence of $\mW^{ }> 0$
plays an important role. Given that a pole lies in the domain
$|p^0_\rmii{E2}| > p$, so that 
$ 
  p^2_{ } - (p^{0}_\rmii{E2} )^2 < 0
$, 
we must have 
$
  1 +  
 \widehat\Sigma^\rmii{E2}_{(p^{0}_\rmiii{E2},p)} > 0 
$. 
At the same time, 
$
 \widehat\Sigma^\rmii{E2}_{(p^{0}_\rmiii{E2},p)} < 0 
$. 
Therefore 
a solution is found in a domain in which the absolute value
$
 | \widehat\Sigma^\rmii{E2}_{(p^{0}_\rmiii{E2},p)} | 
$
is not too large. This leads to 
$
 \lim_{\mW\to 0^+_{ }}^{ } p^0_\rmii{E2} = m^{ }_\rmii{E2} / \sqrt{3}
$, 
so that we are {\em not} on the light-cone, unlike what could 
be assumed by naively setting $\mW^{ }\to 0$ in \eq\nr{zero_E2}. 

In \eq\nr{higgs_2}, the pole terms are evaluated with the four-momentum
$\K - \P = (\omega - p^{0}_{ },-\vec{p})$. The Dirac-$\delta$'s in 
\eqs\nr{jacob_T} and \nr{jacob_E} therefore become 
$\delta(\omega - p^0_{ } - p^0_\alpha)$.
This sets the integration variable $p^0_{ }$
in \eq\nr{higgs_2} to $p^0_{ } = \omega - p^0_\alpha(p)$.

Finally, we need to determine
when the would-be pole crosses into the integration
domain of \eq\nr{higgs_2}. 
For $\omega < \omega^{ }_\rmi{pl}$, 
the minimal value of $p$, 
denoted by $p^{ }_\alpha \equiv p^{ }_{\alpha,\rmi{min}}$, 
is obtained from the condition $p^0_{ } = -p$, 
cf.\ \fig\ref{fig:kinematics}(right). 
Employing $p^0_{ } = \omega - p^0_\alpha$, 
we see that 
$
 \omega + p^{ }_\alpha = p^0_\alpha
$, 
i.e.\ 
\ba
 p^2_\rmii{T2} - (\omega + p^{ }_\rmii{T2})^2_{ } + 
 \Sigma^\rmii{T2}_{(\omega + p^{ }_\rmiii{T2},p^{ }_\rmiii{T2})}
 + \mW^2 & = & 0
 \;, \la{pmin_T2}
 \\[2mm]
 \bigl[
   p^2_\rmii{E2} - (\omega + p^{ }_\rmii{E2} )^2_{ } 
 \bigr]
 \bigl[
  1 +  
 \widehat\Sigma^\rmii{E2}_{(\omega + p^{ }_\rmiii{E2},p^{ }_\rmiii{E2})}
 \bigr]
 + \mW^2 & = & 0
 \;.  \la{pmin_E2}
\ea

Having determined $p^{ }_\alpha$ and $p^0_\alpha$, 
\eq\nr{higgs_2} can be estimated numerically. The value of 
the integrand at $p \gg m^{ }_\rmii{E2}$ can also be found
analytically, and we return to this in \se\ref{ss:soft_2to2}. 

\section{Contribution of $1+n \leftrightarrow 2+n$ processes}
\la{se:1n_2n}

\subsection{Outline}
\la{se:outline_1n_2n}

Apart from the $2\leftrightarrow 2$ scatterings considered in 
\se\ref{se:processes}, the Higgs interaction rate may
also get a contribution from $1\to 2$ decays, $2\to 1$ ``inverse
decays'', or $2\leftrightarrow 3$ processes, amongst them 
``brehmsstrahlung''. 
In contrast to most $2\leftrightarrow 2$ scatterings 
(cf.\ \fig\ref{fig:symmetric}(right)),
$1+n\leftrightarrow 2+n$ processes 
are inelastic.  
The purpose of this section is to 
estimate their magnitude.

The consideration of $1+n\leftrightarrow 2+n$ processes is, however, 
faced with a challenge. This is that the apparently leading reaction, 
$1\leftrightarrow 2$, is severely phase-space constrained, being
open only if the masses of the participating particles 
($m^{ }_1,m^{ }_2,m^{ }_3$) 
satisfy the usual inequalities $m^{ }_3 < |m^{ }_1 - m^{ }_2|$ or 
$m^{ }_3 > m^{ }_1 + m^{ }_2$. Therefore, $n=1$ can give
a larger contribution than $n=0$, even if it nominally 
involves more couplings. This indicates that the naive
perturbative expansion breaks down, and a resummation
over all values of $n$ may be required. We will be 
confronted with this issue in \se\ref{se:higgs_1n_2n}.

Unfortunately, the resummation over $n$ is well understood only
in a particular kinematic domain, namely for hard momenta, $p \gg g T$, 
where it goes 
under the name of Landau-Pomeranchuk-Migdal (LPM) resummation
(cf.,\ e.g.,\ ref.~\cite{lpm} and references therein). In this 
domain, the problem can be mapped onto a dimensionally reduced
effective theory~\cite{sch}, 
and NLO computations become accessible. 
In our situation, when all masses and momenta are soft, 
the lack of a supplementary 
scale hierarchy makes the resummation less
transparent.\footnote{%
 For the computation of transport coefficients, 
 such as viscosities, where the observable can be defined at $k = 0$, 
 the resummation is believed to be under control at leading order, 
 because the contributing quasiparticles carry large momenta  
 (cf.,\ e.g.,\ ref.~\cite{visc_nlo}). 
}

In the present study, we address the $1+n \leftrightarrow 2+n$
processes {\em without} any additional resummation
on top of the HTL one. In other words, 
we consider
$1+n \leftrightarrow 2+n$ processes with either $n=0$ or $n=1$, 
whichever of these channels happens to be open. 
It turns out that in all cases considered, 
only one of them is open at a time. 
We note that the $2\leftrightarrow 3$ processes originate from 
the so-called ``cut--cut contributions''~\cite{dilepton}
of HTL perturbation theory.

\subsection{Confinement phase}
\la{se:confinement_1n_2n}

Starting with the confinement phase, 
and staying in the regime of \eq\nr{regime}, 
a $1\leftrightarrow 2$ process can indeed be found. 
Concretely, it originates
from the $s$-channel part of \eq\nr{conf_general}, 
when this part is used in the 
result in \eq\nr{symm_1}.
Physically, this corresponds to the decay of a ``plasmon'' 
into two $\phi$-particles, or its inverse process. 

By definition, the 
$s$-channel corresponds to the regime $p^0_{ } > p$.
Then the self-energies in \eq\nr{PiT_split} and \nr{hatPiE_split} have no
imaginary parts. Therefore a non-vanishing spectral 
function $\rho^\channel_\P$ can only emerge via \eq\nr{pole}. 
If such a pole exists, 
we call it a plasmon. 
The plasmon dispersion relation, 
which we denote by
$
 p^0_\alpha(p)
$,  
can be found by looking for the zeros of  
\eqs\nr{PiT_split} and \nr{hatPiE_split}~\cite{qed1,qed2,qed3}, 
similarly to \eqs\nr{zero_T2} and \nr{zero_E2}.
This can be done analytically only in specific limits, notably
by expanding either in $p / p^0_{ }$ or $(p^0_{ } - p)/p$, 
\ba 
  p \ll p^0_{ }:
                 && \P^2_{ } + \Pi^\rmii{T2}_\P \approx 
                    p^2 - (p^0_{ })^2_{ } + 
                    \frac{\mE^2}{3}\,\biggl[ 1 
                  + \frac{p^2_{ }}{5 (p^0_{ })^2_{ }} + \ldots \biggr]
  \la{piT2_small} \;, \\[2mm]
                 && \P^2_{ } +  \Pi^\rmii{E2}_\P \approx
                     p^2 - (p^0_{ })^2_{ } + 
                    \frac{\mE^2}{3}\,\biggl[ 1 
                  - \frac{2 p^2_{ }}{5 (p^0_{ })^2_{ }} + \ldots \biggr]
  \la{piE2_small} \;, \\[2mm]
  p^0_{ } - p \ll p:
                 && \P^2_{ } +  \Pi^\rmii{T2}_\P \approx 
                    2 p (p - p^0_{ }) + 
                    \frac{\mE^2}{2}\,\biggl[ 1 
                  + \frac{p^0_{ } - p}{p} 
                    \biggl( 
                       2 - \ln \frac{2p}{p^0_{ } - p}
                    \biggr) + \ldots \biggr]
  \la{piT2_large} \;, \hspace*{5mm} \\[2mm]
                 && \P^2_{ } +  \Pi^\rmii{E2}_\P \approx
                    (p - p^0_{ }) \,\biggl[ 
                    2 p   
                  +   
                    \frac{\mE^2}{p} 
                    \biggl( 
                       2 - \ln \frac{2p}{p^0_{ } - p}
                    \biggr) + \ldots \biggr]
  \;. \la{piE2_large}
\ea
Equations~\nr{piT2_small} and \nr{piE2_small} imply 
that the plasmon energy is $\mE^{ }/\sqrt{3}$ for $p \ll \mE^{ }$. 
In the opposite limit $p \gg \mE^{ }$, it follows from 
\eq\nr{piT2_large} that 
$p^0_\rmii{T2} \approx p + \mE^2 / (4 p)$, and from 
\eq\nr{piE2_large} that $p^0_\rmii{E2}$ is exponentially 
close to the lightcone, 
$
 p^0_\rmii{E2} \approx p + 2 p \exp\bigl(  - 2p^2_{ }/ \mE^2 - 2 \bigr)
$.  

We now observe from \eq\nr{conf_general} that the $2\to 1$ process
takes place if a solution can be found for the equality
$
 p^0_\rmii{E$a$}(p) 
 {=} 
 m^{ }_\phi + \epsilon^{ }_\phi 
$,
where
$
 \epsilon^{ }_\phi \equiv \sqrt{p^2 + m_\phi^2}
$.
For $p\to 0$, the left-hand side reads $\mE^{ }/\sqrt{3}$ and the 
right-hand side $2 m^{ }_\phi$, whereas for $p\to\infty$, the
left-hand side equals $p$ and the right-hand side $p + m^{ }_\phi > p$.
A solution exists
(i.e.\ the curves cross)
if $2m^{ }_\phi < \mE^{ } / \sqrt{3}$, as is indeed
parametrically the case in the regime of \eq\nr{regime}. 

\vspace*{3mm}

To be more precise, combining \eqs\nr{pole}, \nr{symm_1} and 
\nr{conf_general}, working out carefully the sign of the pole term,
and noting that the result originates from the $p^0_{ } > 0$ domain
because of the energy conservation constraint,  
the $s$-channel contribution becomes 
\ba
 && \hspace*{-1.0cm} 
 \im\Delta^{ }_{(-i [m^{ }_\phi + i 0^+_{ }],\vec{0});\phi}
 \; \supset \; - \, 
  \sum_{a=1}^{2} \frac{d^{ }_a g_a^2}{4} 
  \la{conf_1to2_1} \\
 & & \; \times \, 
 \int_\vec{p} 
 \int_{0}^{\infty}\! {\rm d}p^0_{ }\, 
 \frac{p^2_{ } + 4m_\phi^2 - (p^0_{ })^2_{ }}
      {p^2 - (p^0_{ })^2_{ }}
 \,
 \frac{ 
   \nB^{ }(\epsilon^{ }_\phi) - \nB^{ }( p^0_{ } )  
      }
                            {2\epsilon^{ }_\phi}
 \, 
 \pi\, 
 \delta \Bigl( 1 + \re
  \widehat\Sigma^\rmii{E$a$}_\P 
 \Bigr)
 \delta \bigl( m^{ }_\phi + \epsilon^{ }_\phi - p^0_{ } \bigr)
 \;. \nonumber
\ea
In order to integrate over $p^0_{ }$, we make use of a 
variant of \eq\nr{jacob_E},  
\be
 \delta \Bigl( 1 + \re
  \widehat\Sigma^\rmii{E$a$}_\P 
 \Bigr)
 \; \stackrel{p^0_{ }\,>\,0}{\supset} \; 
 \frac{ \delta\bigl(\, p^0_{ } - p^0_\rmiii{E{\it a}}\, \bigr) 
        }{|\partial^{ }_{p^0_{ }} 
              \re  \widehat\Sigma^\rmii{E$a$}_\P 
              |}
 \; = \; 
 \frac{p^0_\rmiii{E{\it a}}\,[(p^0_\rmiii{E{\it a}})^2 - p^2]\,
       \delta\bigl(\, p^0_{ } - p^0_\rmiii{E{\it a}}\, \bigr)}
       {p^2 + m^2_\rmiii{E{\it a}} - (p^0_\rmiii{E{\it a}})^2 }
 \;,
\ee
where $p^0_\rmii{E$a$}$ is the solution of 
\be
 1 + 
 \frac{ m^2_\rmiii{E{\it a}} }{{p}^2}
   \biggl[\,
     1 -     
     \frac{p^0_\rmiii{E{\it a}} }{2p}
   \ln\biggl| \frac{p^0_\rmiii{E{\it a}} + p}
                   {p^0_\rmiii{E{\it a}} - p } \biggr|
   \,\biggr] 
 \;
 \stackrel{p^0_\rmiii{E{\it a}} > p}{=} 
 \;
 0 
 \;. \la{p0_pole}
\ee
Subsequently, the momentum constraint can be expressed as 
\be
 \delta \bigl( m^{ }_\phi + \epsilon^{ }_\phi - p^0_\rmiii{E{\it a}} \bigr)
 \; = \; 
 \frac{\delta( p - p^{ }_\rmiii{E{\it a}}) }
      {|\partial^{ }_p [ \epsilon^{ }_\phi - p^0_\rmiii{E{\it a}} ]|} 
 \;, 
 \quad
 \mbox{with} 
 \quad 
 \frac{{\rm d} p^0_\rmiii{E{\it a}} }{{\rm d}p}
 \; 
  = 
 \; 
 \frac{p^0_\rmiii{E{\it a}}}{p^{ }_\rmiii{E{\it a}}}
 \frac{m^2_\rmiii{E{\it a}}
  + 3 [p^2_\rmiii{E{\it a}} - (p^0_\rmiii{E{\it a}})^2_{ }]}
      {m^2_\rmiii{E{\it a}}
  + p^2_\rmiii{E{\it a}} - (p^0_\rmiii{E{\it a}})^2_{ }}
 \;,
\ee
where $p^{ }_\rmiii{E{\it a}}$ can be found by substituting
the energy conservation constraint into \eq\nr{p0_pole}, 
\be
 1 + 
 \frac{ m^2_\rmiii{E{\it a}} }{{p}^2_\rmiii{E{\it a}}}
   \biggl[\,
     1 -     
     \frac{ m^{ }_\phi + \epsilon^{ }_\phi  }{2p^{ }_\rmiii{E{\it a}}}
   \ln\biggl| \frac{ m^{ }_\phi + \epsilon^{ }_\phi + p^{ }_\rmiii{E{\it a}}}
                   { m^{ }_\phi + \epsilon^{ }_\phi - p^{ }_\rmiii{E{\it a}}
   } \biggr|
   \,\biggr] 
 \;
 = 
 \; 
 0 
 \;, \quad
 \epsilon^{ }_\phi \; = \; \sqrt{p^2_\rmiii{E{\it a}} + m_\phi^2}
 \;. \la{p_pole}
\ee
The Jacobian is a bit intransparent, 
\be
 \frac{1} {\partial^{ }_p [ \epsilon^{ }_\phi - p^0_\rmiii{E{\it a}} ]} 
 =  
 \frac{ 
      p^{ }_\rmiii{E{\it a}} \, 
      \epsilon^{ }_\phi \, 
      [
        m^2_\rmiii{E{\it a}}
      + p^2_\rmiii{E{\it a}}
      - (p^0_\rmiii{E{\it a}})^2_{ } 
      ]
      }{
      p^2_\rmiii{E{\it a}} \, 
      [
        m^2_\rmiii{E{\it a}}
      + p^2_\rmiii{E{\it a}}
      - (p^0_\rmiii{E{\it a}})^2_{ } 
      ]
      -
      p^{0}_\rmiii{E{\it a}} \, 
      \epsilon^{ }_\phi \, 
      \{   
        m^2_\rmiii{E{\it a}}
     + 3 [p^2_\rmiii{E{\it a}}
     - (p^0_\rmiii{E{\it a}})^2_{ }]
      \}
      }
 \;, 
\ee
but it can be simplified by making use of 
the energy conservation constraint 
$
 p^{0}_\rmiii{E{\it a}} = 
 m^{ }_\phi + \epsilon^{ }_\phi 
$.
The final result then takes the form
\be
 \im\Delta^{ }_{(-i [m^{ }_\phi + i 0^+_{ }],\vec{0});\phi}
 \; \supset \; - \, 
  \sum_{a=1}^{2} \frac{d^{ }_a g_a^2}{8\pi} 
 \frac{
       p^{3}_\rmiii{E{\it a}}
       (  \epsilon^{ }_\phi - m^{ }_\phi  )\,
       [
            \nB^{ }(\epsilon^{ }_\phi)
          - \nB^{ }( \epsilon^{ }_\phi + m^{ }_\phi )   
       ]
      }
      {
      2 [  
          2 p^{2}_\rmiii{E{\it a}}
          + 3 m^{ }_\phi (\epsilon^{ }_\phi + m^{ }_\phi)
        ] 
      - 
      m^2_\rmiii{E{\it a}}
      }
 \, 
 \;,   \la{conf_1to2_2} 
\ee
whose numerical evaluation is deferred to \se\ref{ss:soft_1n_to_2n}.

\subsection{Higgs phase}
\la{se:higgs_1n_2n}

We now turn to the Higgs phase. In this case the diagrams
are given by \fig\ref{fig:broken}(c). The contribution of 
$2\leftrightarrow 2$ processes is given by \eq\nr{higgs_2}, 
and this get modified for the $1\leftrightarrow 2$ and
$2\leftrightarrow 3$ processes considered in the present section.

As long as $\mW^{ } > 0$, we can understand the kinematics
of the problem from \fig\ref{fig:kinematics}.
A $1\leftrightarrow 2$ ``pole-pole'' contribution would exist if the 
Higgs decayed into, or were generated from, two plasmons. 
The contribution can be written as 
\ba
 && \hspace*{-1.0cm} 
 \im\Delta^{ }_{(-i [\omega + i 0^+_{ }],\vec{0});h}
 \; \underset{\K \, \equiv \, (\omega + i 0^+_{ },\vec{0})}
             {\overset{\P^2 \,\equiv\, p^2 - (p^0_{ })^2}{\supset}}
 \; 
 - \, g_2^2 
 \int_\vec{p}
 \int_{-\infty}^{\infty}
 \! \frac{{\rm d}p^0_{ } }{\pi}
 \bigl[\,
  1 + \nB^{ }(p^0_{ }) + \nB^{ }(\omega - p^0_{ })
 \,\bigr]
 \nn[3mm] 
 & \times & \biggl\{ \,
 2 \mW^2 \,  
 \im^{ }_\rmii{(pole)} \biggl[\,
 \frac{ 1
      }
      { \P^2 + \Sigma^\rmiii{T2}_{\P} + \mW^2
      }
 \,\biggr]\, 
 \im^{ }_\rmii{(pole)} \biggl[\,
     \frac{1}{ (\K-\P)^2_{ }
      + \Sigma^\rmiii{T2}_{\K-\P} 
      + \mW^2 } 
     \biggr] 
 \nn
 & + & 
 \frac{ [p^2 + p^0_{ }(\omega - p^0_{ })]^2_{ } }{\mW^2} \,
 \im^{ }_\rmii{(pole)} \biggl[\,
 \frac{ 1  + \widehat\Sigma^\rmiii{E2}_{\P}
      }
      { \P^2(1 + \widehat\Sigma^\rmiii{E2}_{\P}) 
      + \mW^2
      }
 \,\biggr]\, 
 \im^{ }_\rmii{(pole)} \biggl[\,
     \frac{1
          + \widehat\Sigma^\rmiii{E2}_{\K-\P} 
          }
          {
         (\K-\P)^2_{ } 
         ( 1 
           + \widehat\Sigma^\rmiii{E2}_{\K-\P} 
         )
             + \mW^2 }  
     \biggr] 
  \,\biggr\}
 \nn[2mm]
 & + & \mbox{($Z$-channel)}
 \;. \hspace*{5mm} \la{higgs_3_pre}
\ea
The poles can be evaluated like in \eqs\nr{jacob_T} and \nr{jacob_E}.
The result involves the constraint
$
 \delta\bigl( p^0_{ } - p^0_\alpha(p) \bigr) 
 \delta\bigl( \omega - p^0_{ } - p^0_\alpha(p) \bigr)
$.
Therefore the reaction is kinematically allowed only if
$
 \omega = m^{ }_h \ge 2 \min\{ p^0_{\alpha}(p) \} = 2 \mE^{ }/ \sqrt{3}
$.
If we are in the domain of \eq\nr{regime}, this is not the case,
and the pole-pole contribution is absent.

Instead, there is a cut-cut contribution where 
the gauge modes are in the $t$-channel, i.e.\ spacelike. 
Making use of the symmetry of the integrand, as well as 
\eq\nr{cut_PiE}, it can be written as 
\ba
 && \hspace*{-1.0cm} 
 \im\Delta^{ }_{(-i [\omega + i 0^+_{ }],\vec{0});h}
 \; \underset{\K \, \equiv \, (\omega,\vec{0})}
             {\overset{\P^2 \,\equiv\, p^2 - (p^0_{ })^2}{\supset}}
 \; 
 - \, 2 g_2^2 \mW^2  
 \int_{\frac{\omega}{2}}^{\infty} \! \frac{{\rm d}p\, p^2}{2\pi^2}
 \int_{\frac{\omega}{2}}^{p}
 \! \frac{{\rm d}p^0_{ } }{\pi}
 \bigl[\,
  1 + \nB^{ }(p^0_{ }) + \nB^{ }(\omega - p^0_{ })
 \,\bigr]
 \nn[3mm] 
 & \times & \biggl\{ \, 
 \frac{2\, \Gamma^\rmii{T2}_{\P} 
      }
      { (\P^2 + \Sigma^\rmii{T2}_{\P} + \mW^2)^2_{ } 
      + (\Gamma^\rmii{T2}_{\P})^2_{ }}
 \frac{\Gamma^\rmii{T2}_{\K-\P} 
      }
      { [ 
         (\K-\P)^2 + \Sigma^\rmii{T2}_{\K-\P} 
       + \mW^2
        ]^2_{ } 
      + [ 
          \Gamma^\rmii{T2}_{\K-\P} 
        ]^2_{ }
      }
 \nn[3mm]
 & + & 
 \frac{[p^2 + p^0_{ }(\omega - p^0_{ })]^2_{ }\,
      \widehat \Gamma^\rmii{E2}_{\P} 
      }
      { [
        \P^2(1 + \widehat\Sigma^\rmii{E2}_{\P}) 
      + \mW^2
        ]^2_{ } 
      + (\P^2 \widehat \Gamma^\rmii{E2}_{\P})^2_{ }
      }
 \frac{\widehat \Gamma^\rmii{E2}_{\K - \P}
      }
      { [
         (\K - \P)^2_{ }
         (1 + \widehat\Sigma^\rmii{E2}_{\K - \P}
         )
      + \mW^2
        ]^2_{ } 
      + [
          (\K - \P)^2_{ }  
          \widehat \Gamma^\rmii{E2}_{\K - \P}
        ]^2_{ }
      }
  \,\biggr\}
 \nn[3mm]
 & + & \mbox{($Z$-channel)}
 \;. \hspace*{5mm} \la{higgs_3}
\ea
Here $\omega = m^{ }_h$.
We return to a numerical evaluation 
in \se\ref{ss:soft_1n_to_2n}. 

\section{Numerical results}
\la{se:soft}

%
\begin{table}[t]

\small{
\begin{center}
\begin{tabular}{ccccccccccc} 
 $ T/\mbox{GeV} $ & 
 $ g_1^2 $ & %
 $ g_2^2 $ & %
 $ g_3^2 $ & %
 $ \lambda $ & %
 $ h_t^2 $ & %
 $ h_b^2 $ & %
 $ m^{ }_\rmii{E1}/ T $ & %
 $ m^{ }_\rmii{E2}/ T $ & %
 $ m^{ }_\rmii{E3}/ T $ &  %
 $ m^{ }_\rmii{F}/ T $   %
 \\[2mm]
 \hline
 \\[-4mm] 
 160    & 
 0.130  & 
 0.408  & 
 1.02  & 
 0.102  & 
 0.733  & 
 0.00484  & 
 0.488  & 
 0.865   & 
 1.45   &
 0.412
   \\[0.5mm]  
 \hline 
\end{tabular} 
\end{center}
}


\caption[a]{\small
 Standard Model values for couplings and Debye masses 
 at $T\simeq 160$~GeV~\cite{mDebye}. For reproducibility, 
 more digits have been shown than are physically accurate. 
 For the quark thermal mass $\mF^{ }$, making an appearance
 in appendix~A, we have employed the 
 leading-order expression from \eq\nr{fermion_regime}.
 }
\label{table:couplings}
\end{table}
%

%
\begin{figure}[t]

 \centerline{
             \epsfysize=7.5cm\epsfbox{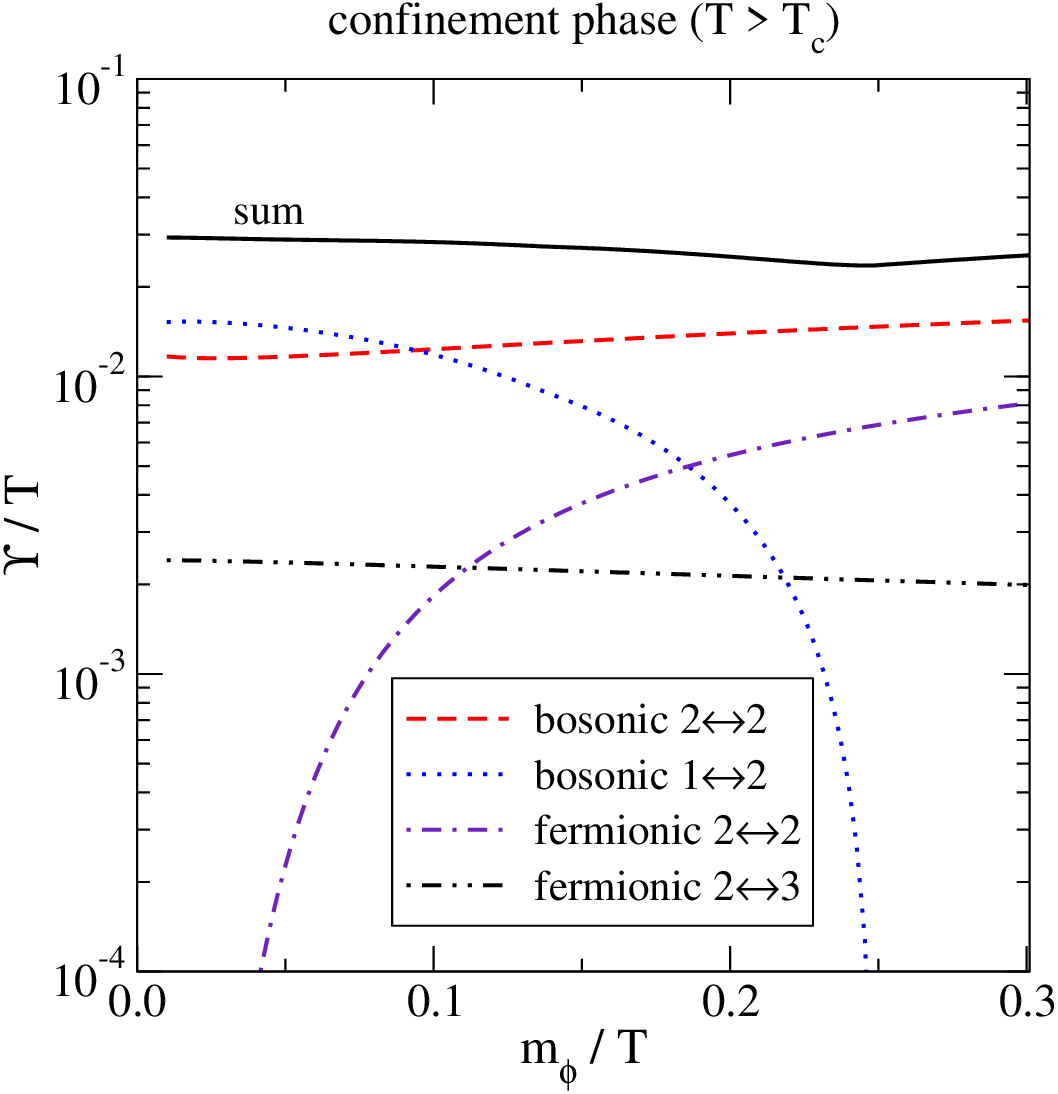}
 }

\vspace*{1mm}

\caption[a]{\small 
 The contributions to $\Upsilon$ 
 in the confinement phase, 
 from 
 \eqs\nr{conf_2_pre}, 
 \nr{bosonic_1to2}, 
 \nr{fermion_3}, 
 and \nr{fermion_4}, 
 after inserting the couplings
 from table~\ref{table:couplings}. 
} 
\la{fig:Upsilon}
\end{figure}
%

\subsection{Outline}

The integral representations in the confinement 
(cf.\ \eqs\nr{conf_1} and \nr{conf_1to2_2})
and Higgs phases (cf.\ \eqs\nr{higgs_2} and \nr{higgs_3}) 
depend on many scales 
($m^{ }_\phi$, $m^{ }_h$, $\mW^{ }$, $\mZ^{ }$, 
$m^{ }_\rmii{E1}$,
$\mE^{ }$, $\pi T$). 
Consequently the functional forms are not transparent, 
and it is unclear if the results remain finite 
if hierarchies emerge between the scales.  
In order to clarify the situation, we consider the soft limit
\be
 m^{ }_\phi\,,\,
 m^{ }_h\,,\,
 \mW^{ }\,,\,
 \mZ^{ }
 \; \ll \; 
 m^{ }_\rmii{E1}\,,\,
 \mE^{ } 
 \; \ll \;
 \pi T 
 \quad \mbox{(soft~limit)}
 \;. \la{regime3}
\ee
The soft regime is physically relevant, as it corresponds to 
being in the confinement phase ($\mW^{ },\mZ^{ }\to 0$) and going close
to the phase transition, in principle even to a 
supercooled domain ($m^{ }_\phi \to 0$). This is precisely
the situation most relevant for bubble nucleation. 
We demonstrate that 
the physics originates
from the exchange of soft momenta, $ p \sim \mEa^{ }$, 
and that \eqs\nr{conf_1} and \nr{higgs_2} 
have the same asymptotics at $ p \gg \mEa^{ }$.

\subsection{$2\leftrightarrow 2$ processes}
\la{ss:soft_2to2}

In the confinement phase expression of \eq\nr{conf_1},
estimating $\epsilon^{ }_\phi \sim p \ll \pi T$, 
the phase space distributions become
\ba
 1 + \nB^{ }(\epsilon^{ }_\phi) + \nB^{ }(m^{ }_\phi - \epsilon^{ }_\phi)
 \stackrel{\epsilon^{ }_\phi \,\ll\, \pi T}{\approx } 
 \frac{m^{ }_\phi T}{\epsilon^{ }_\phi(m^{ }_\phi - \epsilon^{ }_\phi)} 
 \;. \la{soft_limit_conf}
\ea 
Parts of 
$
 \widehat\Gamma^\rmii{E$a$}_{( m^{ }_\phi - \epsilon^{ }_\phi ,p)} 
$
and the prefactor can be combined into
\be
 (\epsilon^{ }_\phi - m^{ }_\phi)
 \biggl[\frac{4 m_\phi^2}
 {p^2 - (m^{ }_\phi - \epsilon^{ }_\phi)^2} + 1 \biggr]
 \; = \; 
 \epsilon^{ }_\phi + m^{ }_\phi
 \;. 
\ee
For the friction coefficient from \eq\nr{Upsilon}
this yields 
\ba
 \Upsilon\bigr|^{T > T^{ }_\rmii{c}}_{\rmi{bosonic}\,2\leftrightarrow 2}
 & \supset & 
 \sum_{a=1}^2
 \frac{ d^{ }_a g_a^2 T}{32\pi}
 \int_0^\infty
 \! {\rm d}p\,
 \Phi^{ }_a(p) \bigr|^{T > T^{ }_\rmii{c}}_{\rmi{bosonic}\,2\leftrightarrow 2}
 \;, \la{conf_2_pre} \\[3mm] 
 \Phi^{ }_a(p) \bigr|^{T > T^{ }_\rmii{c}}_{\rmi{bosonic}\,2\leftrightarrow 2}
 & = & 
 \frac{ 
 \frac{p^3_{ } (\epsilon^{ }_\phi + m^{ }_\phi)\, \mEa^2  }{\epsilon_\phi^2 
 (\epsilon^{ }_\phi - m^{ }_\phi)}
 }
 {
 \Bigl[ p^2 + \mEa^2 
 \Bigl(1 + 
 \frac{\epsilon^{ }_\phi - m^{ }_\phi}{2p}
 \ln \bigl| 
 \frac{m^{ }_\phi - \epsilon^{ }_\phi + p}
      {m^{ }_\phi - \epsilon^{ }_\phi - p}
      \bigr|
 \Bigr) 
 \Bigr]^2_{ }
 + 
 \Bigl[ 
   \frac{\pi (\epsilon^{ }_\phi - m^{ }_\phi) \mEa^2}{2 p} 
 \Bigr]^2_{ }
 }
  \la{conf_2}  
 \;. \hspace*{5mm}
\ea 
The integrand is peaked at $p\sim \mEa^{ }$, 
where its value is $\sim O(1/\mEa^{ })$. 
A numerical evaluation is shown in \fig\ref{fig:Upsilon}.

\vspace*{3mm}

We now demonstrate where to find the same physics on the side
of the Higgs phase. In \eq\nr{higgs_2}, 
we can set $\mW^{ }\to 0$ in the {\sc T}-part, which then drops out. 
It would be tempting to put $\mW^{ }\to 0$ also in the {\sc E}-part. 
However, as discussed below \eq\nr{zero_E2}, the limit needs to 
be taken carefully, in order to pick up the correct pole (otherwise we
find a singular integral). 

Let us write the soft limit of the Higgs phase expression for $\Upsilon$ 
in a form analogous to \eq\nr{conf_2}.
If we add the $Z$-channel, for which $2 g_2^2 \to g_1^2 + g_2^2$
in the limit $\mZ\to 0$, we get
\be
 \Upsilon\bigr|^{T < T^{ }_\rmii{c}}
 _\rmii{$m^{ }_{h},m^{ }_{W} \, \ll \, m^{ }_\rmiii{Ea}$} 
 \;\supset \; 
 \sum_{a=1}^2
 \frac{ d^{ }_a g_a^2 T}{32\pi}
 \int_{p^{ }_\rmiii{Ea}}^\infty
 \! {\rm d}p \, \Phi^{ }_a (p)
 \bigr|^{T < T^{ }_\rmii{c}}
 _\rmii{$m^{ }_{h},m^{ }_{W} \, \ll \, m^{ }_\rmiii{Ea}$} 
 \;, \la{higgs_4_pre}
\ee
where the lower integration bound originates from \eq\nr{pmin_E2}. 
For $p \sim p^{ }_\rmii{Ea} \sim m^{ }_\rmii{Ea}$, the full 
expression from \eqs\nr{higgs_2} and \nr{jacob_E} needs to be 
employed. The pole from \eq\nr{zero_E2} can only be solved for
numerically, and the same holds for $p^{ }_\rmii{Ea}$.\footnote{%
 We remark that at 
 leading-logarithmic order, $p^{ }_\rmii{Ea}$ is given 
 by the analogue of \eq\nr{llog}, just 
 with $m^{ }_\phi\to m^{ }_h$.
}

However, if we consider the contribution from momenta
$p \gg m^{  }_\rmii{Ea}$, life simplifies. 
The self-energy $ \widehat\Sigma^\rmii{E2}_{\P}$ has the prefactor
$m_\rmii{E2}^2 / p^2$ (cf.\ \eq\nr{hatPiE_split}), and therefore
represents a small correction in this domain. Then, from 
\eq\nr{zero_E2}, 
\be
 p^0_\rmii{E2} 
 \stackrel{
   p\, \gg \, m^{ }_\rmiii{Ea}
 }{\approx}
 \pm \,\epsilon^{ }_\rmiii{\it W}
 \;, \quad
 \epsilon^{ }_\rmiii{\it W} \; \equiv \; \sqrt{p^2 + m_\rmiii{\it W}^2}
 \;.
\ee
The Jacobian becomes 
\be
 \im^{ }_\rmii{(pole)} \biggl[\,
     \frac{1 + \widehat\Sigma^\rmiii{E2}_{\P} 
          }
          {
         \P^2_{ } 
         ( 1 
           + \widehat\Sigma^\rmiii{E2}_{\P} 
         )
             + \mW^2 }  
     \biggr] 
 \stackrel{
   p\, \gg \, m^{ }_\rmiii{Ea}
 }{\approx}
 \frac{\pi\,\bigl[ \delta(p^0_{ } - \epsilon^{ }_\rmiii{\it W} )
                 - \delta(p^0_{ } + \epsilon^{ }_\rmiii{\it W} )
                 \,\bigr]}
 {2 \epsilon^{ }_\rmiii{\it W}}
 \;. 
\ee
In the chosen integration domain, recalling that $m^{ }_h - p^0_{ } > 0$
appears as the energy variable in the pole part, we find 
$p^0_{ } \approx m^{ }_h - \epsilon^{ }_\rmiii{\it W}$.
The factors in \eq\nr{higgs_2} can be simplified as 
\ba
 p^2 + p^0_{ }(m^{ }_h - p^0_{ })
 \bigr|^{ }_{p^0_{ } = m^{ }_h - \epsilon^{ }_\rmiii{\it W}}
 & 
 \overset{ p\, \gg \, m^{ }_\rmiii{Ea} }{
 \underset{
   m^{ }_{h},m^{ }_\rmiii{\it W} \, \ll \, m^{ }_\rmiii{Ea}
  }{\approx}}
 & 
 \epsilon^{ }_\rmiii{\it W} m^{ }_h
 \;, \\ 
 p^2 - (p^0_{ })^2_{ } 
 \bigr|^{ }_{p^0_{ } = m^{ }_h - \epsilon^{ }_\rmiii{\it W}}
 &
 \overset{ p\, \gg \, m^{ }_\rmiii{Ea} }{
 \underset{
   m^{ }_{h},m^{ }_\rmiii{\it W} \, \ll \, m^{ }_\rmiii{Ea}
  }{\approx}}
 & 
 2 \epsilon^{ }_\rmiii{\it W} m^{ }_h
 \;. 
\ea
Altogether we find 
\be
 \Phi^{ }_a(p) \bigr|^{T < T^{ }_\rmii{c}}
 _\rmii{$m^{ }_{h},m^{ }_{W} \, \ll \, m^{ }_\rmiii{Ea}$}
 \quad
 \overset{ p\, \gg \, m^{ }_\rmiii{Ea} }{
 \underset{
  }{\approx}}
 \quad
 \frac{p\, m^2_\rmiii{Ea}}{p^4_{ } + \bigl[ 
   \frac{\pi m^2_\rmiii{Ea} }{2}
 \bigr]^2_{ }}
 \;. \la{higgs_4}
\ee
This agrees perfectly with the 
$
 p \gg m^{ }_\rmiii{Ea} \gg m^{ }_\phi
$
limit of \eq\nr{conf_2}. 

If we carry out the full integral, then the value of 
\eq\nr{conf_2_pre} at small $m^{ }_\phi$ and the value of 
\eq\nr{higgs_4_pre} at small $m^{ }_h,\mW^{ }, \mZ^{ }$ 
do not agree completely, with the Higgs phase value
being $\sim 50$\% larger. 
At finite $m^{ }_h$, the Higgs phase value also depends on the ratio
$m^{ }_h/\mW^{ }$, i.e.\ on the scalar self-coupling~$\lambda$. 
However, the limiting value at $m^{ }_h\to 0$ appears to 
be $\lambda$-independent, like in the confinement phase.

\subsection{$1+n \leftrightarrow 2+n$ processes}
\la{ss:soft_1n_to_2n}

For the $1+n \leftrightarrow 2+n$ processes, 
we start with the confinement phase result from 
\eq\nr{conf_1to2_2}, corresponding to $n=0$. 
The transcendental constraint
in \eq\nr{p_pole} cannot be solved analytically, 
but in the soft limit $m^{ }_\phi \ll m^{ }_\rmii{E$a$}$,
a ``leading-logarithmic'' solution works reasonably well, 
\be
 p^{ }_\rmii{E$a$} \; \simeq \;
 m^{ }_\rmii{E$a$} \, 
 \sqrt{ 
       \frac{1}{2} \ln \biggl( 
          \frac{2 m^{ }_\rmiii{E{\it a}}}{m^{ }_\phi}
       \biggr) - 1
      } 
 \;. \la{llog}
\ee 
This shows that $p^{ }_\rmii{E$a$} \sim m^{ }_\rmii{E$a$} \gg m^{ }_\phi$, 
and therefore 
$ 
 \epsilon^{ }_\phi \approx p^{ }_\rmii{E$a$} \gg m^{ }_\phi
$. 
Given that $m^{ }_\rmii{E$a$} \ll \pi T$, it follows that
\be
            \nB^{ }(\epsilon^{ }_\phi)
          - \nB^{ }( \epsilon^{ }_\phi + m^{ }_\phi )
          \; \stackrel{\epsilon^{ }_\phi\,\ll\, \pi T}{\approx} \;
          \frac{m^{ }_\phi T}
               {\epsilon^{ }_\phi(\epsilon^{ }_\phi + m^{ }_\phi)} 
     \;. \la{soft_nB_3}
\ee
All in all this leads to 
\be
 \Upsilon\bigr|^{T > T^{ }_\rmii{c}}_{\rmi{bosonic}\,1\leftrightarrow 2}
 \; \approx \;  
  \sum_{a=1}^{2} \frac{d^{ }_a g_a^2 T}{8\pi} 
 \frac{
       \frac{ 
       p^{3}_\rmiii{E{\it a}}
       (  \epsilon^{ }_\phi - m^{ }_\phi  )\,
       }{
        \epsilon^{ }_\phi\,
        ( \epsilon^{ }_\phi + m^{ }_\phi )   
       }
      }
      {
      2 [  
          2 p^{2}_\rmiii{E{\it a}}
          + 3 m^{ }_\phi (\epsilon^{ }_\phi + m^{ }_\phi)
        ] 
      - 
      m^2_\rmiii{E{\it a}}
      }
 \;.   \la{bosonic_1to2} 
\ee
Fixing $p^{ }_\rmii{E$a$}$ from \eq\nr{p_pole}, 
with the parameters from table~\ref{table:couplings}, 
the result is plotted in \fig\ref{fig:Upsilon}. 

\vspace*{3mm}

In the Higgs phase, 
\eq\nr{higgs_3} may appear to vanish
for $\mW^{ }\to 0$,
due to the overall prefactor $\mW^2$.\footnote{%
 The contribution once again originates from the soft domain, 
 so that we may approximate
 $ 
  1 + \nB^{ }(p^0_{ }) + \nB^{ }(\omega - p^0_{ })
  \approx \omega T / [p^0_{ }(\omega - p^0_{ })]
 $.
 The singularities at $p^0_{ } = 0$ and $p^0_{ } = \omega$ are lifted, 
 because one of the $\widehat\Gamma^\rmiii{E{\it a}}_{ }$'s
 vanishes at the same point. 
 } 
However, the second term in \eq\nr{higgs_3}
becomes IR sensitive, 
as it contains the terms 
$
 1 / \P^4_{ }
$
and 
$
 1 / (\K - \P)^4_{ }
$.
If we send $\mW\to 0^+_{ }$ smoothly, 
keeping the ratio $m^{ }_h/\mW^{ }$ fixed, 
the result remains finite. Its value at $\mW\to 0^+_{ }$
depends strongly on $m^{ }_h/\mW^{ }$, 
i.e.\ on the scalar self-coupling $\lambda$. For the Standard
Model value 
(cf.\ table~\ref{table:couplings}), 
it is about a factor 3 larger than that following 
from \eq\nr{bosonic_1to2}.
In terms of the diagram in \fig\ref{fig:broken}(c), this value
originates from a double IR enhancement of soft scatterings, as 
$(g^6_{ }T/\pi^5_{ }) \times (\pi T)^4_{ } / (g T)^4_{ }
\sim g^2_{ }T/\pi$.

However, we should be cautious about the numerical interpretation of
the Higgs phase expression. The fact that the $1\leftrightarrow 2$
rate vanishes, but that the $2\leftrightarrow 3$ rate, in which the 
the participants of the $1\leftrightarrow 2$ process have been ``dressed''
by their finite interaction rates, is large, suggests a breakdown of the
perturbative expansion. Alas, we are not aware of a clean procedure
to resum the corresponding physics. 

\section{Conclusions and outlook}
\la{se:concl}

Even if the thermodynamic properties 
of an electroweak phase transition 
can be determined with high precision with modern tools
(cf.,\ e.g.,\ ref.~\cite{dralgo} and references therein), the 
understanding of its real-time dynamics, including nucleations
and the subsequent bubble growth and collisions, 
remains partly on a more qualitative level. 
In fact, general doubts about the semi-classical picture have been 
raised recently~(cf.,\ e.g.,\ refs.~\cite{nucl1,nucl15} 
and references therein). 
To make progress, it may become necessary to head in 
the direction of numerical simulations of the full problem
(cf.,\ e.g.,\ ref.~\cite{nucl2} and references therein).

A potential framework for numerical studies is offered by
fluctuating hydrodynamics, with the normal degrees of freedom
supplemented by a scalar order parameter, which satisfies
a Langevin equation. The purpose of our study has
been to estimate the parameters appearing in such a Langevin
equation. We have done this predominantly on the side
of the high-temperature phase, which becomes metastable
as the temperature decreases below $\Tc^{ }$. 

Concretely, we have estimated the width, or interaction rate, 
of a nearly homogeneous Higgs field, 
displaced slightly from its equilibrium value. 
Various contributions to this width
are illustrated in \fig\ref{fig:Upsilon}.
In a concrete BSM model, required for actually generating 
a first-order transition, contributions from other particles
should be added, however we do not expect them to change the
qualitative features that we have observed. 

We remark that the interaction rate 
we have found is quite large, 
of order $g^2 T/\pi$. A significant part 
originates from $2\to 2$ scatterings, which would naively
yield a thermally averaged rate $\sim g^4_{ } T/\pi^3_{ }$. 
However,  the $t$-channel exchange is IR-sensitive, 
regularized by the Debye scale $\sim g^2 T^2$
(such a contribution originates from the longitudinal 
or ``electric'' E-part of the gauge propagator
in \eq\nr{htl_prop}). 
This boosts the naive rate by a factor $\sim (\pi T)^2/(g T)^2$.
Physically, these rapid soft scatterings 
decohere the system. The resulting classical physics
should then lend itself to a Langevin description. 

Even though we have studied quantitatively only
the confinement phase ($T > \Tc^{ }$), 
we have also inspected the formal expressions in the 
Higgs phase ($T < \Tc^{ }$). This has permitted 
for a non-trivial crosscheck of the gauge 
independence of the equilibration rate, and the demonstration
that in both phases it is the E-part polarization state
that plays the decisive role. The results also
demonstrate how $2\leftrightarrow 2$ scatterings are 
rather insensitive to particle masses, 
whereas $1+n\leftrightarrow 2+n$ scatterings depend strongly
on them, particularly in the Higgs phase. 
This suggests the need for a yet-to-be-understood resummation, 
in order to obtain 
quantitative results also in the Higgs phase. 

We end by recalling that 
a similar friction as appears in the Langevin equation,  
is also needed for studies of bubble growth. 
However, in that case gradients are large, 
and momentum exchange plays a role. In the traditional
approach, the momentum exchange originates
from the position dependence of quasiparticle masses
across the wall, $\sim {\rm d} m^2(\bar{v}(z))/{\rm d}z$~\cite{v_wall}.  
Barring IR divergences, this leads to a friction that vanishes in 
the high-temperature phase. Our friction has a non-zero value
in the high-temperature phase, and an even a larger value in the Higgs phase, 
both of which we have treated as homogeneous
(the field perturbations do have ``on-shell'' time dependence). 
In this spirit, we presume that our minimal 
result, depicted in \fig\ref{fig:Upsilon}, 
could represent a lower bound for 
the physical friction affecting bubble growth.

%
%

%
\appendix
\renewcommand{\thesection}{\Alph{section}} 
\renewcommand{\thesubsection}{\Alph{section}.\arabic{subsection}}
\renewcommand{\theequation}{\Alph{section}.\arabic{equation}}

%
\section{Contributions from quark-gluon scatterings}
\la{se:quarks}

%
\begin{figure}[t]
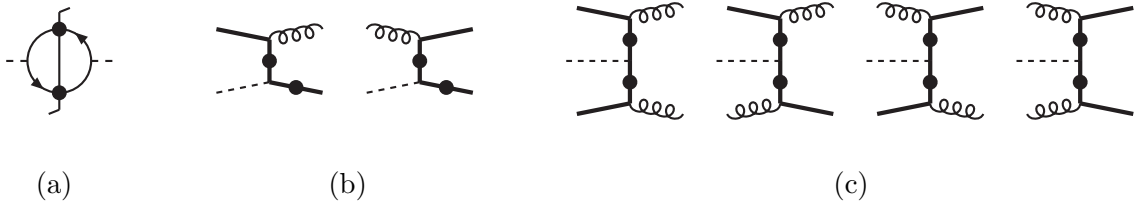


\begin{eqnarray}
&& 
 \hspace*{-0.5cm}
 \cutF   \hspace*{10mm}
 \ampFa  \hspace*{2mm}
 \ampFb  \hspace*{10mm}
 \ampFFa \hspace*{2mm}
 \ampFFb \hspace*{2mm}
 \ampFFc \hspace*{2mm}
 \ampFFd
 \nn[8mm] 
&&            \hspace*{0.0cm} 
   \mbox{(a)} \hspace*{3.4cm}
   \mbox{(b)} \hspace*{6.2cm}
   \mbox{(c)}
 \nonumber
\end{eqnarray}

\vspace*{-3mm}

\caption[a]{\small 
 (a)~the imaginary part of a HTL-resummed Higgs self-energy
 contribution, originating from Yukawa couplings.
 Dashed lines denote scalar fields, 
 arrowed lines quarks, 
 and a blob HTL resummation.
 (b)~``pole-cut'' contributions (or $2\leftrightarrow 2$ scatterings),
 originating from process~(a). 
 These are analyzed in \se\ref{se:fermion_2_2}.
 Thick lines without an arrow represent both quarks and antiquarks, 
 i.e.\ an arrow in either direction, 
 and curly lines stand for gluons. 
 (c)~``cut-cut'' contributions, 
 also originating from process~(a). 
 These are discussed in \se\ref{se:fermion_1n_2n}.
} 
\la{fig:quark}
\end{figure}
%

\subsection{Outline}

In vacuum, the decay 
$h\to b\bar{b}(g)$ belongs to the most important Higgs decay channels 
(here $g$ stands for a gluon). At finite temperature, when 
$m^{ }_h$ is small
(cf.\ \eq\nr{regime}), 
the $1\to 2$ decay or inverse decay $h\leftrightarrow b\bar{b}$ 
is hindered by the quark thermal mass 
$\sim g^{ }_3 T$, 
where $g_3^2 \equiv 4\pi \alphas^{ }$ 
denotes the strong gauge coupling. 
However, the channel can be opened by adding
one or two gluon legs. 
The corresponding diagrams are shown in 
\fig\ref{fig:quark}.

Even if the general structure of the diagrams 
in \fig\ref{fig:quark} looks similar to those 
in \fig\ref{fig:broken}, there is a major difference
in the corresponding interaction rate. 
This becomes clear if we go to 
the soft limit, inspecting momenta $p \sim g^{ }_3 T \ll \pi T$. 
In contrast to  
the Bose enhancement $\rmO(\omega T / p^2_{ })$
seen in \eq\nr{soft_limit_conf}
or \nr{soft_nB_3}, 
the corresponding fermionic expression 
is of $\rmO(\omega/T)$, cf.\ \eq\nr{ferm_weight}.
That is, compared with the bosonic one, 
we expect the quark-gluon contribution to be suppressed by 
$\rmO(p^2/T^2) \sim g_3^2$, supplemented by an associated 
factor $\sim \Nc^{ }/ \pi^2$, 
where $\Nc^{ } = 3$ is the number of colours. 
However, as $g_3^2 \sim 1.0$ at $T\simeq 160$~GeV
(cf.\ table~\ref{table:couplings}),
this may not be a huge suppression numerically.

We compute the contributions of \fig\ref{fig:quark} in the regime
where the Higgs-induced fermion mass, 
$
 m^{ }_\psi \equiv h^{ }_\psi v / \sqrt{2}
$, 
is at most of the same order as the quark thermal mass, 
\be
 m^{ }_\psi \;\lsim\; \mF^{ } \ll \pi T
 \;, \quad 
 \mF^2 = \frac{m_\infty^2}{2} = \frac{g_3^2 \CF^{ }T^2}{8}
 \;, \quad
 \CF^{ }\; \equiv \; \frac{\Nc^2 - 1}{2\Nc^{ }}
 \;, \la{fermion_regime}
\ee
where $\CF^{ }$ is the quadratic Casimir coefficient
of the fundamental representation. 
Here two frequently used thermal mass parametrizations have been displayed, 
$\mF^{ }$ and $m^{ }_\infty$. 
As will be reviewed below, the two masses originate via
the approximate form of a dispersion relation valid for spatial momenta 
$p \ll g^{ }_3 T$ ($\mF^{ }$) or $p \gg g^{ }_3 T$
($m_\infty^{ }$), respectively.  
The numerical value used is given 
in table~\ref{table:couplings}. 

In the regime of \eq\nr{fermion_regime}, 
the HTL self-energies can be evaluated as in a
massless medium~\cite{qed2,qed4}.\footnote{%
  The vacuum mass $m^{ }_\psi$ appears inside the ``hard'' 
  self-energy correction, 
  and if $m^{ }_\psi\sim \pi T$, 
  the medium can no longer be approximated as massless. 
}
If we carry out the analytic continuation 
$i p^{ }_n \to p^0_{ } + i 0^+$, 
then the inverse fermion propagator and 
the fermion propagator can be written, respectively, as
\ba
 \bsl\Delta^{ }_{\P;\psi} & = &
 m^{ }_\psi \,\mathbbm{1}
 + 
 p^{ }_0 \gamma^0_{ } 
 \, \bigl( 1 + c^\rmii{W}_{\P}\bigr)
 + p^{ }_i \gamma^i_{ }
 \, \bigl( 1 + c^\rmii{P}_{\P}\bigr)
 \;, \la{ferm_prop} \\[3mm] 
 \bsl\Delta^{-1}_{\P;\psi} & = & 
 \frac{   
 m^{ }_\psi \,\mathbbm{1}
 - p^{ }_0 \gamma^0_{ } 
 \, \bigl( 1 + c^\rmii{W}_{\P}\bigr)
 - p^{ }_i \gamma^i_{ }
 \, \bigl( 1 + c^\rmii{P}_{\P}\bigr)
 }{ \Delta^{ }_{\P;\psi}  } 
 \;, \\[3mm] 
 \Delta^{ }_{\P;\psi}
 & \equiv &  
 m_\psi^2 
 + p^2_{ }\, \bigl(\, 1 + c^\rmii{P}_\P \,\bigr)^2_{ }
 - (p^0_{ })^2_{ }\, \bigl(\, 1 + c^\rmii{W}_\P \,\bigr)^2_{ }
 \;, 
\ea
where the HTL corrections are similar to those in 
\eqs\nr{PiT_split} and \nr{hatPiE_split}, {\it viz.}  
\ba
 c^\rmii{W}_{\P} & \equiv & 
 \underbrace{ 
  - \frac{\mF^2}{2 p p^0_{ }}
    \ln\biggl| \frac{p^0_{ } + p}{p^0_{ } - p} \biggr|
 }_{ 
 \; \equiv \; r^\rmii{W}_{\P}
 }
 \; + 
 \; 
 \frac{i\pi \mF^2}{2 p p^0_{ }}
 \, 
 \theta(p - |p^0_{ }|)
 \;, \\[2mm] 
 c^\rmii{P}_{\P} & \equiv & 
 \underbrace{ 
  \frac{\mF^2}{p^2}
  \biggl[\, 
  1  - \frac{p^0_{ }}{2 p}
    \ln\biggl| \frac{p^0_{ } + p}{p^0_{ } - p} \biggr|
  \,\biggr]
 }_{ 
 \; \equiv \; r^\rmii{P}_{\P}
 }
 \; + 
 \; 
 \frac{i\pi \mF^2 p^0_{ } }{2 p^3_{ } }
 \, 
 \theta(p - |p^0_{ }|)
 \;.  
\ea
The superscripts W and P remind us that 
in \eq\nr{ferm_prop}, 
these corrections
appear in connection with frequency (``$\omega$'') 
and spatial momentum
(``$p$''), respectively.

With these propagators, 
we have computed the 1-loop HTL-resummed quark-antiquark 
contribution to the Higgs self-energy, and taken its 
imaginary part (cf.\ \fig\ref{fig:quark}).
After setting $\vec{k}\to \vec{0}$, the overall result
takes a form similar to \eq\nr{matsubara_sum},  
\ba
 \im\Delta^{ }_{(-i [\omega + i 0^+_{ }],\vec{0});h}
 & 
 \supset 
 & 
 - \, 2 h^2_{\psi}\Nc^{ } 
 \int_\vec{p}
 \int_{-\infty}^{\infty}
 \! \frac{{\rm d}p^0_{ } }{\pi}
 \bigl[\,
  1 - \nF^{ }(p^0_{ }) - \nF^{ }(\omega - p^0_{ })
 \,\bigr]
 \nn[2mm] 
 & \times & \biggl\{ \, 
 p^0_{ }(\omega - p^0_{ }) \, 
 \im^{ }_\rmii{ } \biggl[\,
     \frac{ 1 + c^\rmii{W}_{\P} }
          {\Delta^{ }_{\P;\psi}} 
     \,\biggr]
 \im^{ }_\rmii{ } \biggl[\,
     \frac{ 1 + c^\rmii{W}_{(\omega-p^0_{ } ,\vec{p})} }
          {\Delta^{ }_{(\omega-p^0_{ } ,\vec{p});\psi}} 
     \,\biggr]
 \nn[2mm]
 & + & 
 p^2_{ } \, 
 \im^{ }_\rmii{ } \biggl[\,
     \frac{ 1 + c^\rmii{P}_{\P} }
          {\Delta^{ }_{\P;\psi}} 
     \,\biggr]
 \im^{ }_\rmii{ } \biggl[\,
     \frac{ 1 + c^\rmii{P}_{(\omega-p^0_{ } ,\vec{p})} }
          {\Delta^{ }_{(\omega-p^0_{ } ,\vec{p});\psi}} 
     \,\biggr]
 \nn[2mm]
 & - & 
 m^2_{\psi} \, 
 \im^{ }_\rmii{ } \biggl[\,
     \frac{ 1 }
          {\Delta^{ }_{\P;\psi}} 
     \,\biggr]
 \im^{ }_\rmii{ } \biggl[\,
     \frac{ 1 }
          {\Delta^{ }_{(\omega-p^0_{ } ,\vec{p});\psi}} 
     \,\biggr]
  \,\biggr\}
 \;. \hspace*{5mm} \la{fermion_full} 
\ea
Here the frequencies should be understood to contain
the imaginary part $ + i 0^+_{ } $. 

In the remainder of this section, we go to the confinement phase or, 
in the language of \se\ref{se:soft}, to the soft limit, setting
$m^{ }_\psi\to 0$. This simplifies \eq\nr{fermion_full} considerably, 
as the last term drops out, and the inverse propagator factorizes,
\ba
 \lim_{m^{ }_\psi\to 0}
 \Delta^{ }_{\P;\psi} & = & 
 p^2_{ } \bigl(\, 1 + c^\rmii{P}_\P \,\bigr)^2_{ }
 - (p^0_{ })^2_{ } \bigl(\, 1 + c^\rmii{W}_\P \,\bigr)^2_{ }
 \nn 
 & \equiv & 
 \underbrace{
 \bigl[
   p\, \bigl(\, 1 + c^\rmii{P}_\P \,\bigr)
 - p^0_{ }\, \bigl(\, 1 + c^\rmii{W}_\P \,\bigr)
 \bigr]
 }_{ \;\equiv\; \Delta^-_\P }
 \underbrace{
 \bigl[
   p\, \bigl(\, 1 + c^\rmii{P}_\P \,\bigr)
 + p^0_{ }\, \bigl(\, 1 + c^\rmii{W}_\P \,\bigr)
 \bigr]
 }_{ \;\equiv\; \Delta^+_\P }
 \;. \la{Delta_m_p}
\ea
We can then write the curly brackets in \eq\nr{fermion_full} as
\ba
 \{ ... \} & \stackrel{m^{ }_\psi\,\to\,0}{=} & 
 \im\Biggl[ \frac{ p^0_{ }\, \bigl(\, 1 + c^\rmii{W}_\P \,\bigr) }
                 { \Delta^-_\P\, \Delta^+_\P }
    \Biggr]
 \im\Biggl[ \frac{ (\omega - p^0_{ })\, 
                   \bigl(\, 1 +
                 c^\rmii{W}_{(\omega-p^0_{ } ,\vec{p})} \,\bigr) }
                 { \Delta^-_{(\omega-p^0_{ } ,\vec{p})}
                   \,
                   \Delta^+_{(\omega-p^0_{ } ,\vec{p})} 
                  }
     \Biggr]
 \nn 
 &  & \; + \,  
 \im\Biggl[ \frac{ p\, \bigl(\, 1 + c^\rmii{P}_\P \,\bigr) }
                 { \Delta^-_\P\, \Delta^+_\P }
    \Biggr]
 \im\Biggl[ \frac{ p\, 
                   \bigl(\, 1 +
                 c^\rmii{P}_{(\omega-p^0_{ } ,\vec{p})} \,\bigr) }
                 { \Delta^-_{(\omega-p^0_{ } ,\vec{p})}
                   \,
                   \Delta^+_{(\omega-p^0_{ } ,\vec{p})} 
                  }
     \Biggr]
 \nn 
 & = & 
 \frac{1}{4} \Biggl\{ 
 \im\Biggl[ \frac{ 1 } 
                 { \Delta^-_\P } 
          - \frac{ 1 }
                 { \Delta^+_\P }
    \Biggr]
 \im\Biggl[ \frac{ 1 }
                 {
                   \Delta^-_{(\omega-p^0_{ } ,\vec{p})} 
                 }
          - \frac{ 1 }
                  {
                   \Delta^+_{(\omega-p^0_{ } ,\vec{p})} 
                  }
     \Biggr]
 \nn 
 &  & \; + \,  
 \im\Biggl[ \frac{ 1 } 
                 { \Delta^-_\P } 
          + \frac{ 1 }
                 { \Delta^+_\P }
    \Biggr]
 \im\Biggl[ \frac{ 1 }
                 {
                   \Delta^-_{(\omega-p^0_{ } ,\vec{p})} 
                 }
          + \frac{ 1 }
                  {
                   \Delta^+_{(\omega-p^0_{ } ,\vec{p})} 
                  }
     \Biggr]
 \Biggr\}
 \nn 
 & = & 
 \frac{1}{2} \Biggl\{ 
 \im\Biggl[ \frac{ 1 } 
                 { \Delta^-_\P } 
    \Biggr]
 \im\Biggl[ \frac{ 1 }
                 {
                   \Delta^-_{(\omega-p^0_{ } ,\vec{p})} 
                 }
     \Biggr]
 \; + \;  
 \im\Biggl[ \frac{ 1 }
                 { \Delta^+_\P }
    \Biggr]
 \im\Biggl[ \frac{ 1 }
                  {
                   \Delta^+_{(\omega-p^0_{ } ,\vec{p})} 
                  }
     \Biggr]
 \Biggr\}
 \;. \la{partial_fract}
\ea
Adopting this limit, 
in the following we evaluate \eq\nr{fermion_full} 
for $2\leftrightarrow 2$ and 
$1+n\leftrightarrow 2+n$ processes separately. 

\subsection{$2\leftrightarrow 2$ processes}
\la{se:fermion_2_2}

We first consider $2\leftrightarrow 2$ processes, 
corresponding to \fig\ref{fig:quark}(b). 
Having in mind \fig\ref{fig:kinematics}, 
they originate from pole-cut contributions. 
Making use of the symmetry of the integrand, 
we may restrict to the lower pole-cut branch.
{}From \eqs\nr{fermion_full} and \nr{partial_fract}, we then  
obtain an expression analogous to that in \eq\nr{higgs_2}, 
\ba
 && \hspace*{-1.0cm} 
 \im\Delta^{ }_{(-i [\omega + i 0^+_{ }],\vec{0});h}
 \;  
 \stackrel{m^{ }_\psi \to 0 }{\supset} 
 \;
 - \, 2 h^2_{\psi}\Nc^{ } 
 \int_\vec{p}
 \int_{-p}^{\min(p,\omega-p)}
 \! \frac{{\rm d}p^0_{ } }{\pi}
 \bigl[\,
  1 - \nF^{ }(p^0_{ }) - \nF^{ }(\omega - p^0_{ })
 \,\bigr]
 \nn[2mm] 
 & \times & \biggl\{ \, 
 \im^{ }_\rmii{(cut)} \biggl[\,
     \frac{ 1 }
          {\Delta^-_{\P}} 
     \,\biggr]
 \im^{ }_\rmii{(pole)} \biggl[\,
     \frac{ 1 }
          {\Delta^-_{(\omega-p^0_{ } ,\vec{p})}} 
     \,\biggr]
 \; + \; 
 \im^{ }_\rmii{(cut)} \biggl[\,
     \frac{ 1 }
          {\Delta^+_{\P}} 
     \,\biggr]
 \im^{ }_\rmii{(pole)} \biggl[\,
     \frac{ 1 }
          {\Delta^+_{(\omega-p^0_{ } ,\vec{p})}} 
     \,\biggr]
  \,\biggr\}
 \;. \hspace*{5mm} \la{fermion_2} 
\ea
The cut part originates
from the spacelike (or $t$-channel) domain $|p^0_{ }| < p$, 
whereas the pole part comes 
from the timelike (or $s$-channel) domain $|\omega - p^0_{ }| > p$. 
The physical interpretations 
of the quasiparticles corresponding to the 
poles of $\Delta^\pm_\P$ have been discussed in ref.~\cite{pls}. 

To find the poles, we need to search for the zeros of 
$
 \Delta^\pm_{(\omega-p^0_{ },\vec{p})}
$.
In analogy with \eqs\nr{piT2_small}--\nr{piE2_large}, 
we may first inspect the expansions
\ba 
  p \ll p^0_{ }:
                 && \Delta^-_\P \approx 
                    p - p^0_{ }
                  + \frac{\mF^2}{ (p^0_{ })^2_{ } }\,\biggl[ 
                     p^0_{ } - \frac{p}{3} 
                   + \ldots \biggr]
  \la{Delta_m_small} \;, \\[2mm]
                 && \Delta^+_\P \approx
                    p + p^0_{ }
                   - \frac{\mF^2}{ (p^0_{ })^2_{ } }\,\biggl[ 
                     p^0_{ } + \frac{p}{3} 
                    + \ldots \biggr]
  \la{Delta_p_small} \;, \\[2mm]
  p^0_{ } - p \ll p:
                 && \Delta^-_\P \approx 
                    p - p^0_{ }
                  + \frac{\mF^2}{p}\,\biggl[ 1 
                  - \frac{p^0_{ } - p}{2p}  
                      \ln \frac{2p}{p^0_{ } - p}
                     + \ldots \biggr]
  \la{Delta_m_large} \;, \\[2mm]
                 && \Delta^+_\P \approx
                    2 p 
                  + \frac{\mF^2}{p}\,\biggl[ 1  
                  - 
                      \ln \frac{2p}{p^0_{ } - p}
                     + \ldots \biggr]
  \;. \la{Delta_p_large}
\ea
Equation~\nr{Delta_m_small} implies that 
the plasmon energy following from $\Delta^-_\P$
is $p^0_{-} \approx \mF^{ } + p/3$ for $p \ll \mF^{ }$, 
and \eq\nr{Delta_p_small}
that the plasmon energy following from $\Delta^+_\P$ is 
$p^0_{+} \approx \mF^{ } - p/3$ in the same domain.
In the opposite limit $p \gg \mF^{ }$, 
\eq\nr{Delta_m_large} leads to the plasmon energy  
$p^0_{-} \approx p + \mF^2 / p $, whereas 
\eq\nr{Delta_p_large} requires $p$ to be 
exponentially close to the lightcone, 
$
 p^0_{+} \approx p + 2 p \exp\bigl(  - 2p^2_{ }/ \mF^2 -1 \bigr)
$.
In \eq\nr{fermion_2}, the frequency appears as $\omega - p^0_{ }$, 
and it is positive, so we can directly use these forms with 
the substitution $p^0_{\alpha}\to \omega - p^0_{\alpha}$, 
$\alpha = \pm$. 

Returning to \fig\ref{fig:kinematics}(right), we observe 
that if we take $\omega \ll \omega^{ }_\rmi{pl} = \mF^{ }$, 
in accordance with \eq\nr{regime3}, 
then the minimal value of $p$ is pushed to the right, 
$p^{ }_\rmi{min} \gsim \mF^{ }$. To be concrete, from 
\eq\nr{Delta_m_large}, we obtain 
\be
 \im^{ }_\rmii{(pole)} \biggl[\,
     \frac{ 1 }
          {\Delta^-_{(\omega-p^0_{ } ,\vec{p})}} 
     \,\biggr]
 \approx 
 \im \biggl[\,
     \frac{ 1 } 
          { p - \omega + p^0_{ } + \frac{ \mFmini^2 }{ p } - i 0^+_{ }}
     \,\biggr]
 = 
 \pi \, \delta\biggl(
     p - \omega + p^0_{ } + \frac{ \mF^2 }{ p }
   \biggr)
 \;. 
\ee
This crosses inside the allowed domain, $p^0_{ } > -p$, when
\be
 p^0_{ } \approx 
 \omega - p - \frac{\mF^2}{p} > -p 
 \; \Longrightarrow \; 
 p > \frac{\mF^2}{\omega} 
 \stackrel{\mFmini^{ } \gg\, \omega}{\gg} \mF^{ }
 \;. 
\ee
In the cut part, we can then write 
\be
 \Delta^-_\P  = 
 (p - p^0_{ })\, 
 \biggl[\, 1 + \frac{\mF^2}{2p^2_{ }} 
 \ln\biggl| \frac{p^0_{ } + p}{p^0_{ } - p} \biggr| \,\biggr]
 + \frac{\mF^2}{p} - \frac{i\pi \mF^2}{2p^2_{ }}\, (p - p^0_{ })
 \; 
  \overset{ p^0_{ } \,\approx\, -p }
  { \underset{ p \,\gg\, \mFmini^{ } }{ \approx} } 
 \;  
 2 p - \frac{i \pi \mF^2}{p}
 \;, \la{Delta_m_full}
\ee
so that 
\be
  \im^{ }_\rmii{(cut)} \biggl[\,
     \frac{ 1 }
          {\Delta^-_{\P}} 
     \,\biggr]
  \; \approx \;
  \frac{\frac{\pi \mFmini^2}{p}}{4p^2_{ } 
      + \bigl( \frac{\pi \mFmini^2}{p}\bigr)^2_{ }}
  \; \stackrel{ p \, \gg \, \mFmini^{ }}{\approx} \; 
  \frac{\pi \mF^2}{4 p^3_{ }}
  \;.
\ee
Furthermore, 
\be
 1 - \nF^{ }(p^0_{ }) - \nF^{ }(\omega - p^0_{ })
 \; \stackrel{\omega \,\ll\, p^0_{ } }{\approx} \; 
 \frac{\omega}{T} \, \nF^{ }(p^0_{ }) \, 
                     \bigl[ 1 - \nF^{ }(p^0_{ }) \bigr]
 \; \stackrel{ p^0_{ } \, \approx \, -p }{\approx} \; 
 \frac{\omega}{T} \, \nF^{ }(p) \, 
                     \bigl[ 1 - \nF^{ }(p) \bigr]
 \;. 
\ee
Therefore the first part of \eq\nr{fermion_2} yields 
\be
 \Upsilon \; = \; 
 - \frac{ 
 \im\Delta^{ }_{(-i [\omega + i 0^+_{ }],\vec{0});h}
   }{\omega} 
 \;  
 \stackrel{m^{ }_\psi \to 0 }{\supset} 
 \;
 \frac{h^2_{\psi}\Nc^{ } \mF^2 }{4 \pi T} 
 \int_{\frac{\mFmini^2}{\omega}}^{\infty} \! \frac{{\rm d}p}{p}
 \, \nF^{ }(p) \, 
                     \bigl[ 1 - \nF^{ }(p) \bigr]
 \;. \hspace*{5mm} \la{fermion_3} 
\ee
It is appropriate to stress that the Fermi distributions are necessary
for making the integral convergent, i.e.\ that momenta $p\sim \pi T$ play 
a role. Therefore the HTL computation we have presented, 
valid for momenta 
$p\sim \mF^{ }$, is not the full result. Nevertheless, 
it does represent the qualitative behaviour, 
as the sensitivity to the scales $p\sim \pi T$ is only logarithmic. 

We still need to estimate the contribution from the second
term in \eq\nr{fermion_2}. We write the pole part as 
\be
 \im^{ }_\rmii{(pole)} \biggl[\,
     \frac{ 1 }
          {\Delta^+_{(\omega-p^0_{ } ,\vec{p})}} 
     \,\biggr]
 \; = \; 
 - \pi\, \delta\bigl(\,\re  \Delta^+_{(\omega-p^0_{ } ,\vec{p})} \,\bigr) 
 \; = \; 
 - \frac{\pi \, \delta(p^0_{ } - p^0_+ ) 
        }{|\partial^{ }_{p^0_{ }} 
              \re  \Delta^+_{(\omega-p^0_{ } ,\vec{p})}
              |}
 \;. 
\ee
Here, 
\ba
 \re \Delta^+_{(\omega-p^0_{ } ,\vec{p})}  & = &  
 (p + \omega - p^0_{ })\, 
 \biggl[\, 1 - \frac{\mF^2}{2p^2_{ }} 
 \ln\biggl| \frac{\omega - p^0_{ } + p}{\omega - p^0_{ } - p} \biggr| \,\biggr]
 + \frac{\mF^2}{p} 
 \;, \la{Delta_p_re} 
 \\ 
 \partial^{ }_{p^0_{ }}\re \Delta^+_{(\omega-p^0_{ } ,\vec{p})}  
 & 
 \underset{ 
           \;\;\; p^0_{ } \,=\, p^0_+
          }{ = } 
 &  
 - \frac{2 \mF^2}{(\omega - p^0_+)^2 - p^2}
 \;. 
\ea
As mentioned below \eq\nr{Delta_p_large}, 
for $p \gg \mF^{ }$
the pole is exponentially
close to the light cone, 
$
 \omega - p^0_+ 
 \approx
 p + 2 p \exp\bigl(  - 2p^2_{ }/ \mF^2 -1 \bigr)
$.
Therefore, 
$
 1 / | \partial^{ }_{p^0_{ }}\re \Delta^+_{(\omega-p^0_{ } ,\vec{p})} | 
$
is exponentially small, and the second term 
in \eq\nr{fermion_2} gives a subdominant contribution. 

To summarize, the order of magnitude 
of fermionic $2\leftrightarrow 2$ scatterings
is given by \eq\nr{fermion_3}. This result is illustrated numerically 
in \fig\ref{fig:Upsilon}.

\subsection{$1+n\leftrightarrow 2+n$ processes}
\la{se:fermion_1n_2n}

Finally we turn to $2\leftrightarrow 3$ processes, 
illustrated in \fig\ref{fig:quark}(c). 
In the language of \fig\ref{fig:kinematics}, 
they originate from cut-cut contributions. 
Making use of the symmetry of the integrand,
we now restrict to the upper half of the cut-cut domain, 
like in \eq\nr{higgs_3}.  
{}From \eqs\nr{fermion_full} and \nr{partial_fract}, we then  
obtain a variant of \eq\nr{fermion_2}, 
\ba
 && \hspace*{-1.0cm} 
 \im\Delta^{ }_{(-i [\omega + i 0^+_{ }],\vec{0});h}
 \;  
 \stackrel{m^{ }_\psi \to 0 }{\supset} 
 \;
 - \, 2 h^2_{\psi}\Nc^{ } 
 \int_{\frac{\omega}{2}}^{\infty} \! \frac{{\rm d}p\, p^2}{2\pi^2}
 \int_{\frac{\omega}{2}}^{p}
 \! \frac{{\rm d}p^0_{ } }{\pi}
 \bigl[\,
  1 - \nF^{ }(p^0_{ }) - \nF^{ }(\omega - p^0_{ })
 \,\bigr]
 \nn[2mm] 
 & \times & \biggl\{ \, 
 \im^{ }_\rmii{(cut)} \biggl[\,
     \frac{ 1 }
          {\Delta^-_{\P}} 
     \,\biggr]
 \im^{ }_\rmii{(cut)} \biggl[\,
     \frac{ 1 }
          {\Delta^-_{(\omega-p^0_{ } ,\vec{p})}} 
     \,\biggr]
 \; + \; 
 \im^{ }_\rmii{(cut)} \biggl[\,
     \frac{ 1 }
          {\Delta^+_{\P}} 
     \,\biggr]
 \im^{ }_\rmii{(cut)} \biggl[\,
     \frac{ 1 }
          {\Delta^+_{(\omega-p^0_{ } ,\vec{p})}} 
     \,\biggr]
  \,\biggr\}
 \;. \hspace*{5mm} \la{fermion_4} 
\ea
Here, apart from $\Delta^-_{\P}$ in \eq\nr{Delta_m_full}, we need
\be
 \Delta^+_\P  = 
 (p + p^0_{ })\, 
 \biggl[\, 1 - \frac{\mF^2}{2p^2_{ }} 
 \ln\biggl| \frac{p^0_{ } + p}{p^0_{ } - p} \biggr| \,\biggr]
 + \frac{\mF^2}{p} + \frac{i\pi \mF^2}{2p^2_{ }}\, (p + p^0_{ })
 \;. \la{Delta_p_full}
\ee

Let us estimate the magnitude of the integrand at large momenta. 
If $p^0_{ } \sim p \gg \mF^{ }$, then 
\be
 \im^{ }_\rmii{(cut)} \biggl[\,
     \frac{ 1 }
          {\Delta^+_{\P}} 
     \,\biggr]
 \; \sim \;  
 \im^{ }_\rmii{(cut)} \biggl[\,
     \frac{ 1 }
          {\Delta^+_{(\omega-p^0_{ } ,\vec{p})}} 
     \,\biggr]
 \; \stackrel{p^0_{ }\, \sim\, p\,\gg\, \mF^{ }}{\sim} 
 \; \frac{\pi \mF^2}{p^3}
 \;.
\ee
The contribution from $1/\Delta^{-}_{ }$ is 
of the same order of magnitude. 
Therefore the integrand decreases at large momenta fast enough to be 
convergent, even without the Fermi distributions. In other words, the
integration domain is centered 
at $|p^0_{ }| \sim p \ll \pi T$. In this domain, 
$\nF^{ }(p^0_{ }) \approx \frac{1}{2} \bigl( 1 - \frac{p^0_{ }}{2T} \bigr)$, 
and 
\be
  1 - \nF^{ }(p^0_{ }) - \nF^{ }(\omega - p^0_{ })
 \approx 
 \frac{\omega}{4T}
 \;. \la{ferm_weight} 
\ee
A numerical illustration is shown in \fig\ref{fig:Upsilon}. 
The approach to the limit $\omega\to 0$ 
is smooth, unlike that following from \eq\nr{fermion_3}, which becomes
exponentially small if $\omega \ll \mF^2 / T$.

\small{
%

}


\begin{thebibliography}{99}

\bibitem{lisa}
  C.~Caprini \textit{et al.} [LISA Cosmology Working Group],
  {\it Gravitational waves from first-order phase transitions in LISA:
  reconstruction pipeline and physics interpretation,}
  2403.03723.

\bibitem{meg}
  M.E.~Carrington,
  {\it Effective potential at finite temperature in the Standard Model,}
  Phys.\ Rev.\ D {45} (1992) 2933.

\bibitem{gw}
  J.~Ghiglieri and U.A.~Wiedemann,
  {\it Thermal width of the Higgs boson in hot QCD matter,}
  Phys.\ Rev.\ D {99} (2019) 054002
  [1901.04503].

\bibitem{linde}
  A.D.~Linde,
  {\it Infrared problem in thermodynamics of the Yang-Mills gas,}
  Phys.\ Lett.\ {B 96} (1980) 289.

\bibitem{gv2}
  A.~Gynther and M.~Veps\"al\"ainen,
  {\it Pressure of the Standard Model near the electroweak phase transition,}
  JHEP {03} (2006) 011
  [hep-ph/0512177].

\bibitem{lm}
  M.~Laine and M.~Meyer,
  {\it Standard Model thermodynamics across the electroweak crossover,}
  JCAP {07} (2015) 035
  [1503.04935].

\bibitem{gt}
  O.~Gould and T.V.I.~Tenkanen,
  {\it Perturbative effective field theory expansions
  for cosmological phase transitions,}
  JHEP {01} (2024) 048
  [2309.01672].

\bibitem{htl_old}
  T.S.~Bir\'o and M.H.~Thoma,
  {\it Damping rate and Lyapunov exponent of a Higgs field at high
  temperature,}
  Phys.\ Rev.\ D {54} (1996) 3465
  [hep-ph/9603339].

\bibitem{bary1}
  V.A.~Rubakov and M.E.~Shaposhnikov,
  {\it Electroweak baryon number non-conservation in the early Universe 
  and in high-energy collisions,}
  Usp.\ Fiz.\ Nauk {166} (1996) 493
  [Phys.\ Usp.\  {39} (1996) 461]
  [hep-ph/9603208].

\bibitem{bary2}
  D.~B\"odeker and W.~Buchm\"uller,
  {\it Baryogenesis from the weak scale to the grand unification scale,}
  Rev.\ Mod.\ Phys.\ {93} (2021) 035004
  [2009.07294].
  
\bibitem{asy}
  P.B.~Arnold, D.~Son and L.G.~Yaffe,
  {\it The hot baryon violation rate is $\rmO(\alpha_\rmi{w}^5 T^4_{ })$,}
  Phys.\ Rev.\ D {55} (1997) 6264
  [hep-ph/9609481].

\bibitem{db}
  D.~B\"odeker, 
  {\it Effective dynamics of soft non-Abelian gauge fields at finite
  temperature,}
  Phys.\ Lett.\ B {426} (1998) 351
  [hep-ph/9801430].
  
\bibitem{sph_gdm}
  G.D.~Moore,
  {\it Sphaleron rate in the symmetric electroweak phase,}
  Phys.\ Rev.\ D {62} (2000) 085011
  [hep-ph/0001216].

\bibitem{sph_sm}
  M.~D'Onofrio, K.~Rummukainen and A.~Tranberg,
  {\it Sphaleron Rate in the Minimal Standard Model,}
  Phys.\ Rev.\ Lett.\ {113} (2014) 141602
  [1404.3565].

\bibitem{bbl1}
  G.D.~Moore and K.~Rummukainen,
  {\it Electroweak bubble nucleation, nonperturbatively,}
  Phys.\ Rev.\ D {63} (2001) 045002
  [hep-ph/0009132].

\bibitem{bbl2}
  O.~Gould, S.~G\"uyer and K.~Rummukainen,
  {\it First-order electroweak phase transitions: A nonperturbative update,}
  Phys.\ Rev.\ D {106} (2022) 114507
  [2205.07238].

\bibitem{lgv1}
  M.G.~Alford, H.~Feldman and M.~Gleiser,
  {\it Thermal nucleation of kink-antikink pairs,}
  Phys.\ Rev.\ Lett.\ {68} (1992) 1645.

\bibitem{lgv2}
  S.~Bors\'anyi, A.~Patk\'os, J.~Polonyi and Z.~Sz\'ep,
  {\it Fate of the classical false vacuum,}
  Phys.\ Rev.\ D {62} (2000) 085013
  [hep-th/0004059].

\bibitem{hydro}
  G.~Jackson and M.~Laine,
  {\it Hydrodynamic fluctuations from a weakly coupled scalar field,}
  Eur.\ Phys.\ J.\ C {78} (2018) 304
  [1803.01871].

\bibitem{ae}
  A.~Ekstedt,
  {\it Bubble nucleation to all orders,}
  JHEP {08} (2022) 115
  [2201.07331].

\bibitem{ikkl}
  J.~Ignatius, K.~Kajantie, H.~Kurki-Suonio and M.~Laine,
  {\it Growth of bubbles in cosmological phase transitions,}
  Phys.\ Rev.\ D {49} (1994) 3854
  [astro-ph/9309059].

\bibitem{htl1}
 J.~Frenkel and J.C.~Taylor,
 {\it Hard thermal QCD, forward scattering and effective actions,}
 Nucl.\ Phys.\ B {374} (1992) 156.
 
\bibitem{htl2}
 E.~Braaten and R.D.~Pisarski,
 {\it Simple effective Lagrangian for hard thermal loops,}
 Phys.\ Rev.\ D {45} (1992) 1827. 

\bibitem{htl3}
  E.~Braaten and R.D.~Pisarski,
  {\it Soft amplitudes in hot gauge theories: A general analysis,}
  Nucl.\ Phys.\ B {337} (1990) 569.
  
\bibitem{htl4}
  J.C.~Taylor and S.M.H.~Wong,
  {\it The effective action of hard thermal loops in QCD,}
  Nucl.\ Phys.\ B {346} (1990) 115.

\bibitem{qed1}
  V.P.~Silin, 
  {\it On the electromagnetic properties of a relativistic plasma,}
  Sov.\ Phys.\ JETP {11} (1960) 1136
  [Zh.\ Eksp.\ Teor.\ Fiz.\ {38} (1960) 1577].
  
\bibitem{qed2}
  V.V.~Klimov,
  {\it Collective Excitations in a Hot Quark Gluon Plasma,}
  Sov.\ Phys.\ JETP {55} (1982) 199
  [Zh.\ Eksp.\ Teor.\ Fiz.\  {82} (1982) 336].
  
\bibitem{qed3}
  H.A.~Weldon,
  {\it Covariant calculations at finite temperature: The relativistic plasma,}
  Phys.\ Rev.\ D {26} (1982) 1394.

\bibitem{dilepton}
  E.~Braaten, R.D.~Pisarski and T.C.~Yuan,
  {\it Production of soft dileptons in the quark-gluon plasma,}
  Phys.\ Rev.\ Lett.\ {64} (1990) 2242

\bibitem{broken}
  J.~Ghiglieri and M.~Laine,
  {\it Neutrino dynamics below the electroweak crossover,}
  JCAP {07} (2016) 015
  [1605.07720].

\bibitem{lpm}
  J.~Ghiglieri and M.~Laine,
  {\it Smooth interpolation between thermal Born and LPM rates,}
  JHEP {01} (2022) 173
  [2110.07149].

\bibitem{sch}
  S.~Caron-Huot,
  {\it $O(g)$ plasma effects in jet quenching,}
  Phys.\ Rev.\ D {79} (2009) 065039
  [0811.1603].

\bibitem{visc_nlo}
  J.~Ghiglieri, G.D.~Moore and D.~Teaney,
  {\it QCD shear viscosity at (almost) NLO,}
  JHEP {03} (2018) 179
  [1802.09535].

\bibitem{mDebye}
  M.~Laine, P.~Schicho and Y.~Schr\"oder,
  {\it A QCD Debye mass in a broad temperature range,}
  Phys.\ Rev.\ D {101} (2020) 023532
  [1911.09123].

\bibitem{dralgo}
  A.~Ekstedt, P.~Schicho and T.V.I.~Tenkanen,
  {\it DRalgo: A package for effective field theory approach
  for thermal phase transitions,}
  Comput.\ Phys.\ Commun.\ {288} (2023) 108725
  [2205.08815].

\bibitem{nucl1}
  M.~Hindmarsh {\it et al.} [QUEST-DMC collaboration],
  {\it A-B transition in superfluid ${}^3_{ }$He
  and cosmological phase transitions,}
  2401.07878.

\bibitem{nucl15}
  O.~Gould, A.~Kormu and D.J.~Weir,
  {\it A nonperturbative test of nucleation calculations
  for strong phase transitions,}
  2404.01876.

\bibitem{nucl2}
  D.~P\^\i{}rvu, M.C.~Johnson and S.~Sibiryakov,
  {\it Bubble velocities and oscillon precursors
  in first order phase transitions,}
  2312.13364.

\bibitem{v_wall}
  G.D.~Moore and T.~Prokopec,
  {\it How fast can the wall move? A study of the electroweak phase 
  transition dynamics,}
  Phys.\ Rev.\ D {52} (1995) 7182
  [hep-ph/9506475].

\bibitem{qed4}
  H.A.~Weldon,
  {\it Effective fermion masses of order $gT$ 
  in high-temperature gauge theories
  with exact chiral invariance,}
  Phys.\ Rev.\ D {26} (1982) 2789.

\bibitem{pls}
  H.A.~Weldon,
  {\it Dynamical holes in the quark-gluon plasma}, 
  Phys.\ Rev.\  D {40} (1989) 2410.

\end{thebibliography}
\end{document}